\documentclass[a4paper,12pt]{article}

\usepackage{authblk}
\usepackage{graphicx} 
\usepackage{subfigure}
\usepackage{float}
\usepackage{multirow}

\usepackage[symbol]{footmisc}

\title{\textbf{Semi-metals as potential thermoelectric materials: case of HgTe}}
\author[1]{Maxime Markov}
\author[2]{Xixiao Hu}
\author[1]{Han-Chun Liu}
\author[3]{Naiming Liu}
\author[2]{Joseph Poon}
\author[2,3,4]{Keivan Esfarjani}
\author[1,3]{Mona Zebarjadi\footnote{Corresponding author. \textit{E-mail:} m.zebarjadi@virginia.edu}}

\affil[1]{Department of Electrical and Computer Engineering, University of Virginia, Charlottesville, Virginia 22904, USA}
\affil[2]{Department of Physics, University of Virginia, Charlottesville, Virginia 22904, USA}
\affil[3]{Department of Materials Science and Engineering, University of Virginia, Charlottesville, Virginia 22904, USA}
\affil[4]{Department of Mechanical and Aerospace Engineering, University of Virginia, Charlottesville, Virginia 22904, USA}

\date{}

\begin{document}

\maketitle

\begin{abstract}
The best thermoelectric materials are believed to be heavily doped semiconductors. The presence of a bandgap is assumed to be essential to achieve large thermoelectric power factor
and figure of merit. In this work, we study HgTe as an example semimetal with competitive thermoelectric properties. We employ \textit{ab initio} calculations with hybrid exchange-correlation
functional to accurately describe the electronic band structure in conjunction with the Boltzmann Transport theory to investigate the electronic transport properties. We show that
intrinsic HgTe, a semimetal with large disparity in its electron and hole masses, has a high thermoelectric power factor that is comparable to the best known thermoelectric materials.
We also calculate the lattice thermal conductivity using first principles calculations and evaluate the overall figure of merit. Finally, we prepare semi-metallic HgTe samples and 
we characterize their transport properties. We show that our theoretical calculations agree well with the experimental data.\\
\vspace{2cm}
\end{abstract}

\section{Introduction}
Since its discovery in 1821, thermoelectricity remains in the center of interests of the scientific community. Thermoelectric effect (Seebeck effect) refers to direct conversion of thermal to electrical energy in solids and can be used for power generation and waste heat recovery.~\cite{Riffat:2003, Zebarjadi:2012,Champier:2017,Zhao:2014}. Despite their clean, environmentally friendly and reliable performances, thermoelectric modules are only used in niche applications such as in powering space probes. The main obstacle preventing 
thermoelectric technology to be widely used on a mass market today is its relatively low efficiency~\cite{Davis:2006}. 

The thermoelectric efficiency is an increasing function of the material's dimensionless figure of merit $ZT = \frac{S^2\sigma}{\kappa}T$
where $S$ is the Seebeck coefficient, $\sigma$ is the electrical conductivity, $\kappa$ is the thermal conductivity, and $T$ is the absolute temperature. The first two quantities can be combined together into the thermoelectric power factor $P_{F} = S^2\sigma$ describing electronic transport, in contrast to the thermal conductivity, $\kappa$, related to thermal
transport. The power factor is often used as a guide to preselect the class of potential thermoelectric materials. Indeed, metals have highest electrical conductivity but 
suffer from a low Seebeck coefficient. The reason for their low Seebeck coefficient is the symmetry of the density of states around the chemical potential. The number of hot
electrons above the chemical potential in a metal is roughly the same as the number of cold empty states below the chemical potential. As a result under a temperature gradient, 
the number of electrons diffusing from the hot side to the cold side, is approximately equal to the number of cold electrons diffusing from the cold side to the hot side. The same
problem does not exist in semiconductors due to the presence of a band gap allowing only one type of the carriers to diffuse. Typical Seebeck coefficient of semiconductors is two 
orders of magnitude larger than metals. Ioffe first noticed this advantage of semiconductors ~\cite{Ioffe:1932} and paved the way for many successful demonstration of doped semiconductors with high ZT 
values. Later, several research groups including Chasmar \& Stratton~\cite{Chasmar:1959} and Sofo \& Mahan~\cite{Sofo:1994} studied the effect of band gap on thermoelectric properties of materials employing 
two-band toy models for electronic structure and reached the conclusion that best thermoelectrics must have band gap greater than at least 6$k_B$T. 
Today, this criteria has become a golden rule and heavily doped semiconductors are the main focus of the thermoelectric society~\cite{Dehkordi:2015}. While 
opening a band gap is a proven way of increasing the Seebeck coefficient, in this article we show that to have a large Seebeck coefficient, a band gap is not a must. What needed is 
an asymmetric density of states which could be achieved also in semi-metals with slight overlap of electrons and holes bands but with large asymmetry in the electron and hole 
effective masses. 

We turn our attention to semi-metallic HgTe whose properties are in the transition region between semiconductors and metals. HgTe has a very high electron/hole effective 
mass ratio $m_e/m_h \simeq 0.1$~\cite{Berger:1997} which results in large values of the Seebeck coefficient between -90 $\mu V/K$~\cite{Whitsett:1972} and -135 
$\mu V/K$~\cite{Dziuba:1964} at room temperatures which is similar to the Seebeck coefficient of heavily doped semiconductors with a bandgap. 
The carrier concentration of intrinsic HgTe is only $10^{16}-10^{17} cm^{-3}$ which is much smaller than a metal or a typical good heavily-doped semiconductor thermoelectric. 
However, the large electron mobility in HgTe ($\mu> 10^4 cm^2/V.s $)~\cite{Berger:1997} makes up for its low carrier concentration and as a result, the
electrical conductivity of an intrinsic sample is relatively large and is about $\sigma = 1700$ $S/cm$~\cite{Dziuba:1964,Whitsett:1972} at room temperatures.  The large electron 
mobility is partly due to the small effective mass of the electrons and partly because of the absence of dopants. The mobility of a heavily doped semiconductor is limited by ionized 
impurity scattering which is not the case in an intrinsic semi-metal. The experiment reveals that intrinsic HgTe is a high power factor material with $P_F = 14-31$ $\mu$W cm$^{-1}$ 
K$^{-2}$ at T = 300 K~\cite{Dziuba:1964,Whitsett:1972} that is comparable to well-known thermoelectric materials such as SnSe ($P_{F} \simeq 10$ $\mu$W cm$^{-1}$ K$^{-2}$), PbTe$_{1-x}$Se$_{x}$ ($P_{F} \simeq 25$ 
$\mu$W cm$^{-1}$ K$^{-2}$) and Bi$_2$Te$_3$ ($P_{F} \simeq 50$ $\mu$W cm$^{-1}$ K$^{-2}$) at their ZT maximum~\cite{Beekman:2015}. Apart from having a good electrical transport 
properties, mercury telluride is a good thermal insulator with $\kappa = 2.1$ W/mK~\cite{Whitsett:1972,Slack:1972} at T = 300 K. The overall ZT of intrinsic single crystal without
any optimization is between 0.4 to 0.5 and is comparable with most good thermoelectric materials at room temperature. 

The most recent theoretical study of HgTe concludes that semimetallic HgTe (zinc-blende phase) is a poor thermoelectric material with room temperature ZT values close to zero in intrinsic
samples~\cite{Chen:2008} and emphasize the superior thermoelectric performance of a high pressure semiconducting cinnabar phase.~\cite{Chen:2008,Ouyang:2015} 
However, these studies rely on a standard GGA-PBE exchange-correlation functional to describe the electronic structure of a semimetallic HgTe which fails to reproduce the asymmetry in the density of states near the Fermi 
level. Moreover, the use of the same constant relaxation time at different doping concentrations results in an erroneous conclusion that the electrical conductivity always grows with 
the increase of doping. On the contrary, the experimental data shows the drastic decrease of the electrical conductivity with doping in $p$-type samples of HgTe~\cite{Whitsett:1972}.  

In this work, we perform a combined theoretical and experimental study of thermoelectric properties of HgTe at high temperatures. To address the above mentioned issues, we employ 
\textit{ab initio} calculations with hybrid exchange-correlation functional in conjunction with the Boltzmann Transport theory with energy dependent relaxation times obtained from
the fitting of experimental electrical conductivity. We do not attempt to optimize the thermoelectric properties of HgTe using nanostructuring, alloying or slight doping. Instead,
we attempt to develop a platform based on first principles calculations to study its transport properties and to make a case for semi-metals as potential candidates for thermoelectric applications.  

\section{Results and discussion}

\subsection{Electrical transport}

The electronic band structure of zinc-blende HgTe has been extensively studied over the past decade.~\cite{Feng:2010,Svane:2011,Sakuma:2011,Nicklas:2011} It has been shown that \textit{ab initio}
calculations with standard LDA and GGA exchange-correlation functionals can not accurately describe the band structure of HgTe. To achieve a good agreement with experiment, one must perform either GW calculations~\cite{Svane:2011,Sakuma:2011}
or use a hybrid functional~\cite{Feng:2010,Nicklas:2011} where a portion of exact Fock exchange interaction is introduced into a standard exchange-correlation functional. 

\begin{figure}[h]
\centering
\includegraphics[width=0.9\linewidth]{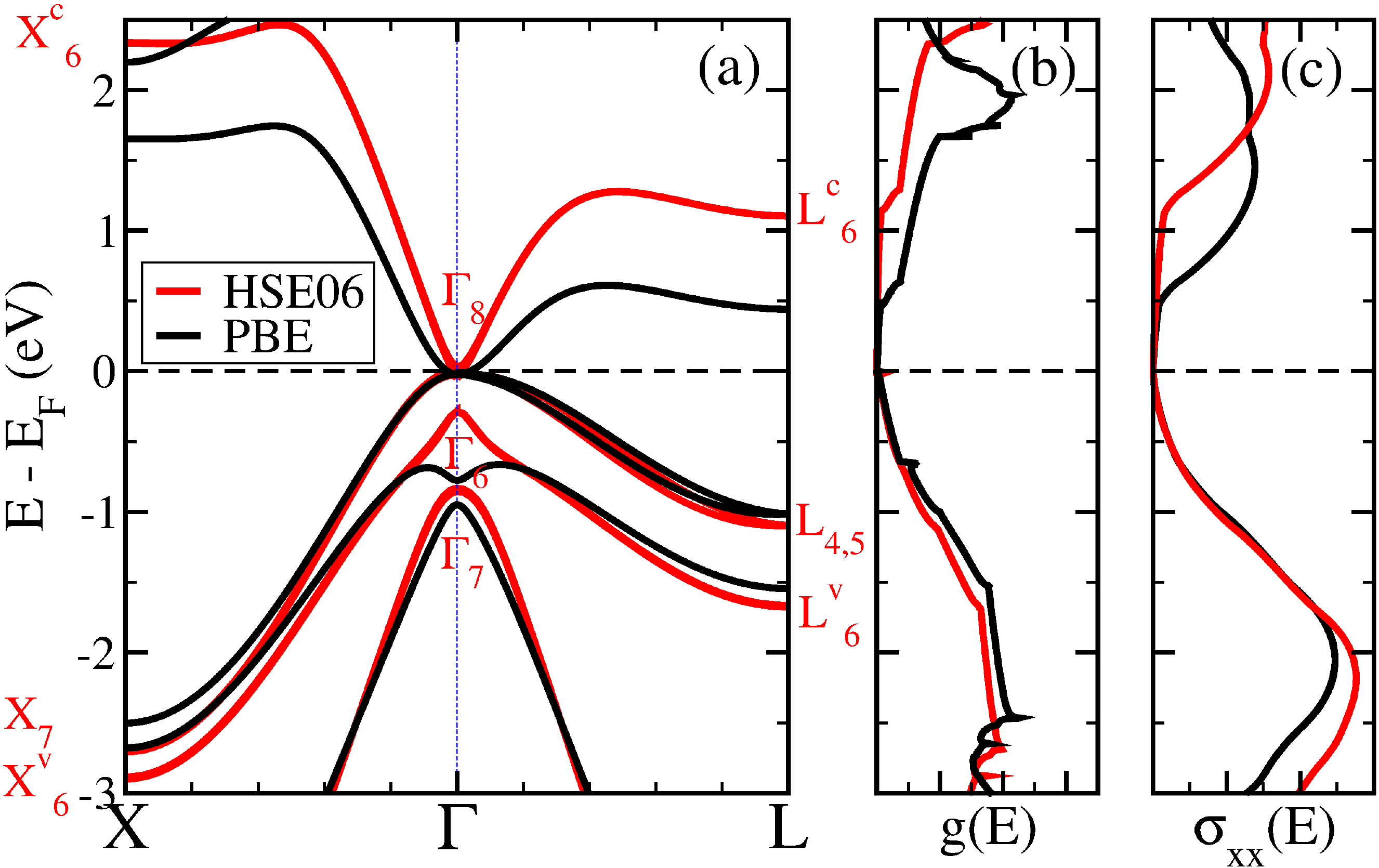}
\caption{Electronic band structure (panel a), density of states $g(E)$ (panel b) and differential conductivity $\sigma_{xx}(E)$ (panel c) calculated using PBE (black curves) and 
HSE06 (red curves) exchange-correlation functionals. Energy levels from the latter calculation are labeled according to their symmetries.}
\label{fig:bandstructure}
\end{figure}

\begin{table}[t]
\centering
\begin{tabular}{lccc}
\hline
  &  GGA-PBE & HSE06 & Expt.\\
\hline
$E_{\Gamma}$ = E($\Gamma_6$) - E($\Gamma_8$)      &  -0.93 & -0.27 & -0.29~\cite{Orlowski:2000},-0.30~\cite{Chadi:1972}\\
$\Delta_{\Gamma}$ = E($\Gamma_8$) - E($\Gamma_7$) &   0.76 &  0.89 &  0.91~\cite{Orlowski:2000}\\
$E_{L}$ = E($L_{6}^{c}$) - E($L_{4,5}$)           &   1.45 &  2.19 & 2.25~\cite{Chadi:1972}\\
$\Delta_{L}$ = E($L_{4,5}$) - E($L_{6}^{v}$)      &   0.54 &  0.56 & 0.62~\cite{Chadi:1972}, 0.75~\cite{Scouler:1964}\\
$E_{X}$ = E($X_{6}^{c}$) - E($X_{7}$)             &   4.15 &  5.02 & 5.00~\cite{Scouler:1964}\\
$\Delta_{X}$ = E($X_{7}$) - E($X_{6}^{v}$)        &   0.19 &  0.22 & 0.1-0.2~\cite{Scouler:1964}\\
\hline
\end{tabular}
\caption{\label{tab:bandstructure} Energy band edges, $E$, and spin-orbit splittings, $\Delta$, at $\Gamma$, $L$ and X high symmetry points calculated with the GGA-PBE
and hybrid-HSE06 functionals. Experimental results from the literature are shown.}
\end{table}

\begin{figure*}[t]
\hfill
\subfigure{\centering\includegraphics[width=0.49\linewidth]{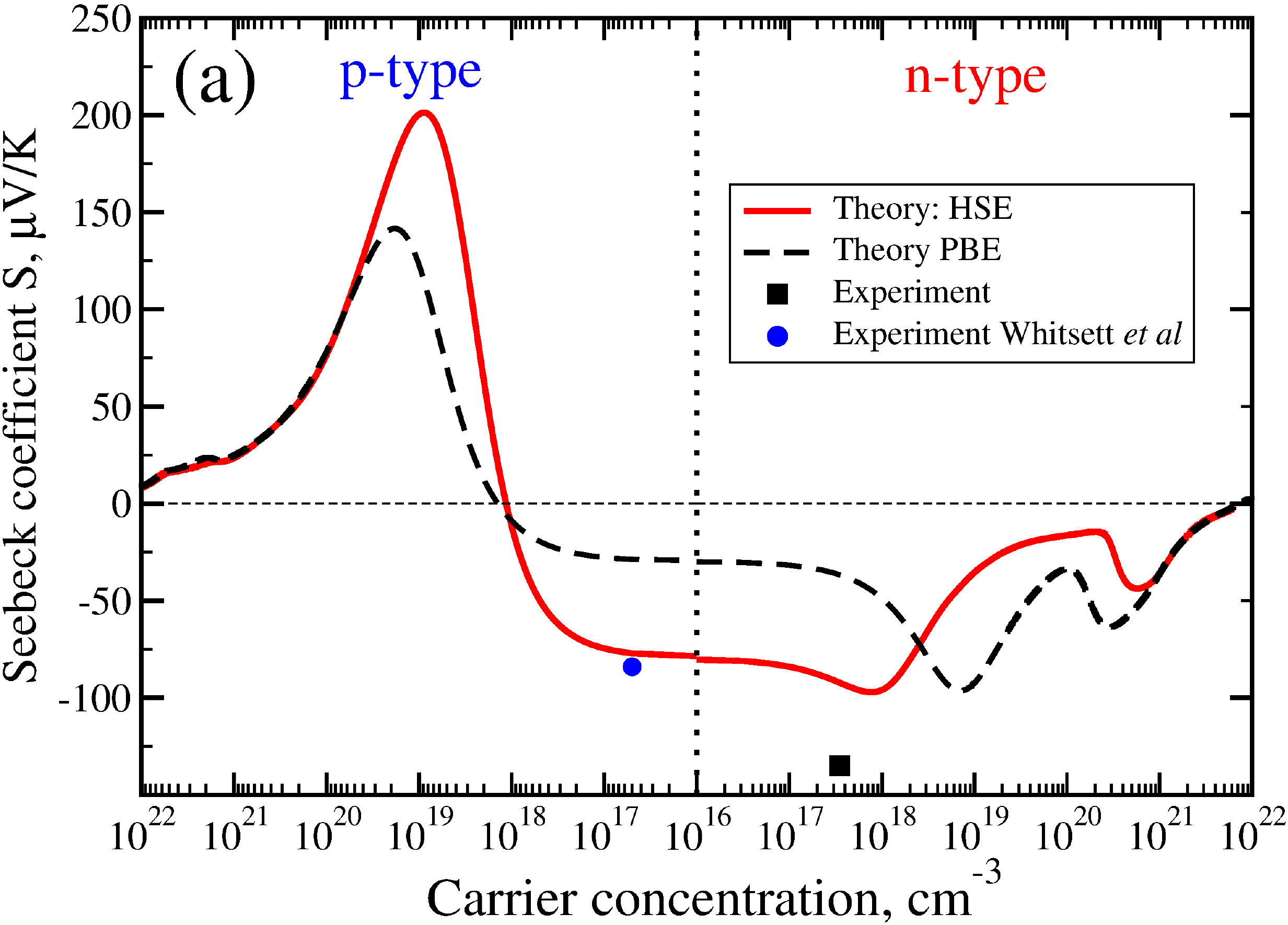}}
\hfill
\subfigure{\centering\includegraphics[width=0.49\linewidth]{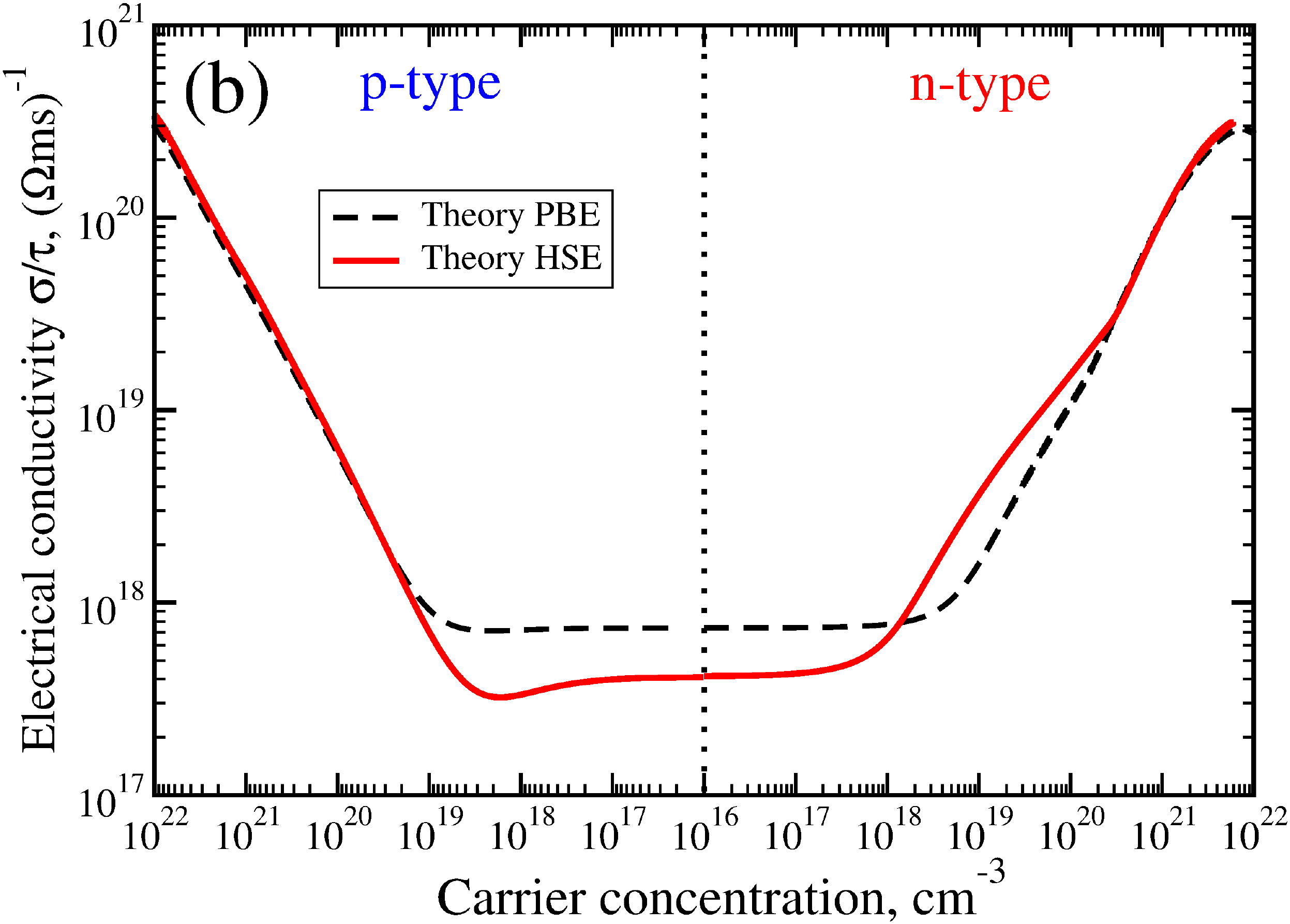}}
\hfill
\caption{The Seebeck coefficient (panel (a)) and the electrical conductivity (panel (b)) as a function of carrier concentration for \textit{p}-type and \textit{n}-type samples at T = 290 K calculated with GGA-PBE (black dashed line) and hybrid-HSE06 (solid red line)
functionals. Experimental data from Whitsett \textit{et al.}~\cite{Whitsett:1972} is shown by a blue circle and experimental data measured in this work is shown by black square.}
\label{fig:elect_transport_300K}
\end{figure*}

\begin{figure*}[t]
\hfill
\subfigure{\includegraphics[width=0.49\linewidth]{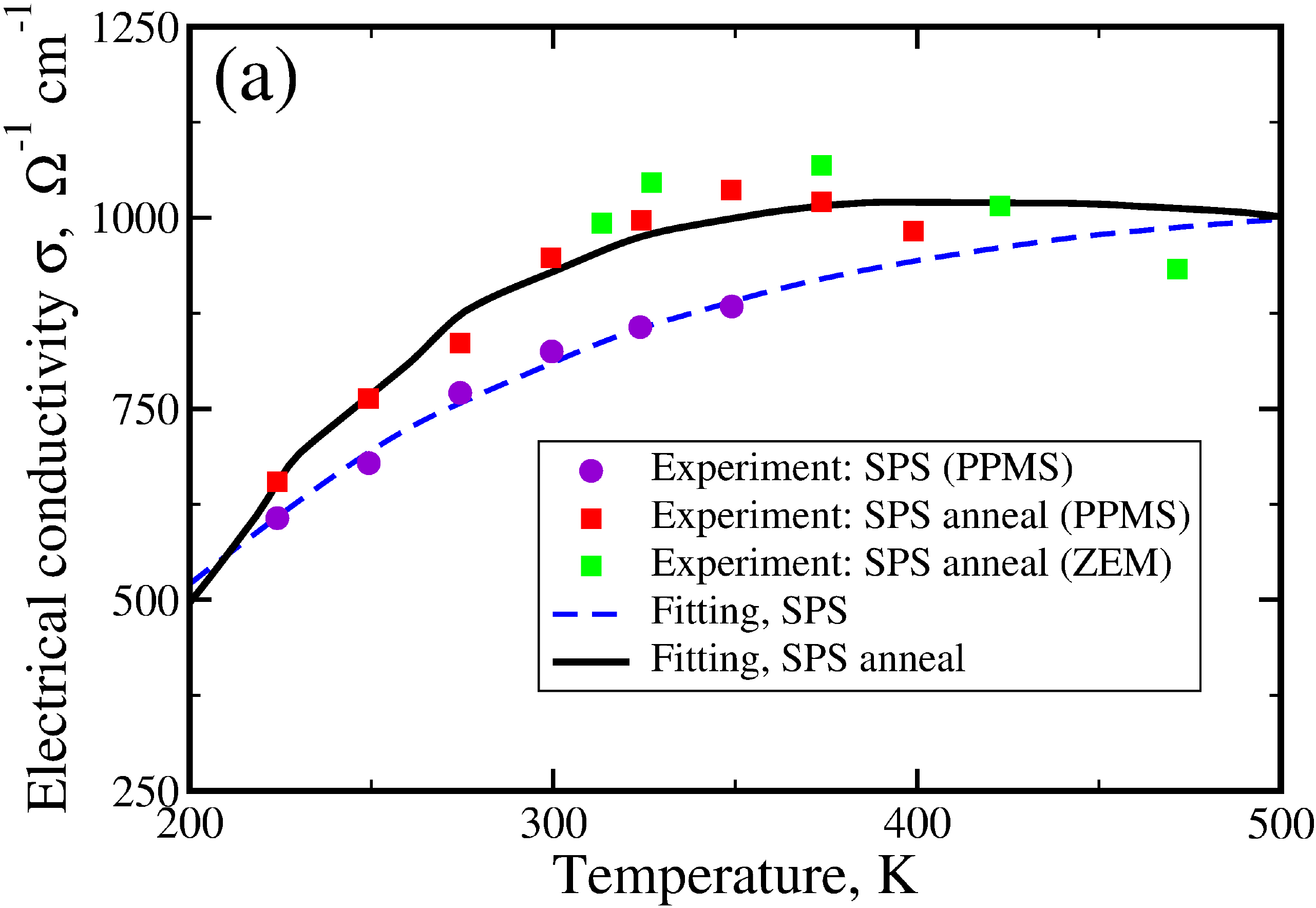}}
\hfill
\subfigure{\includegraphics[width=0.49\linewidth]{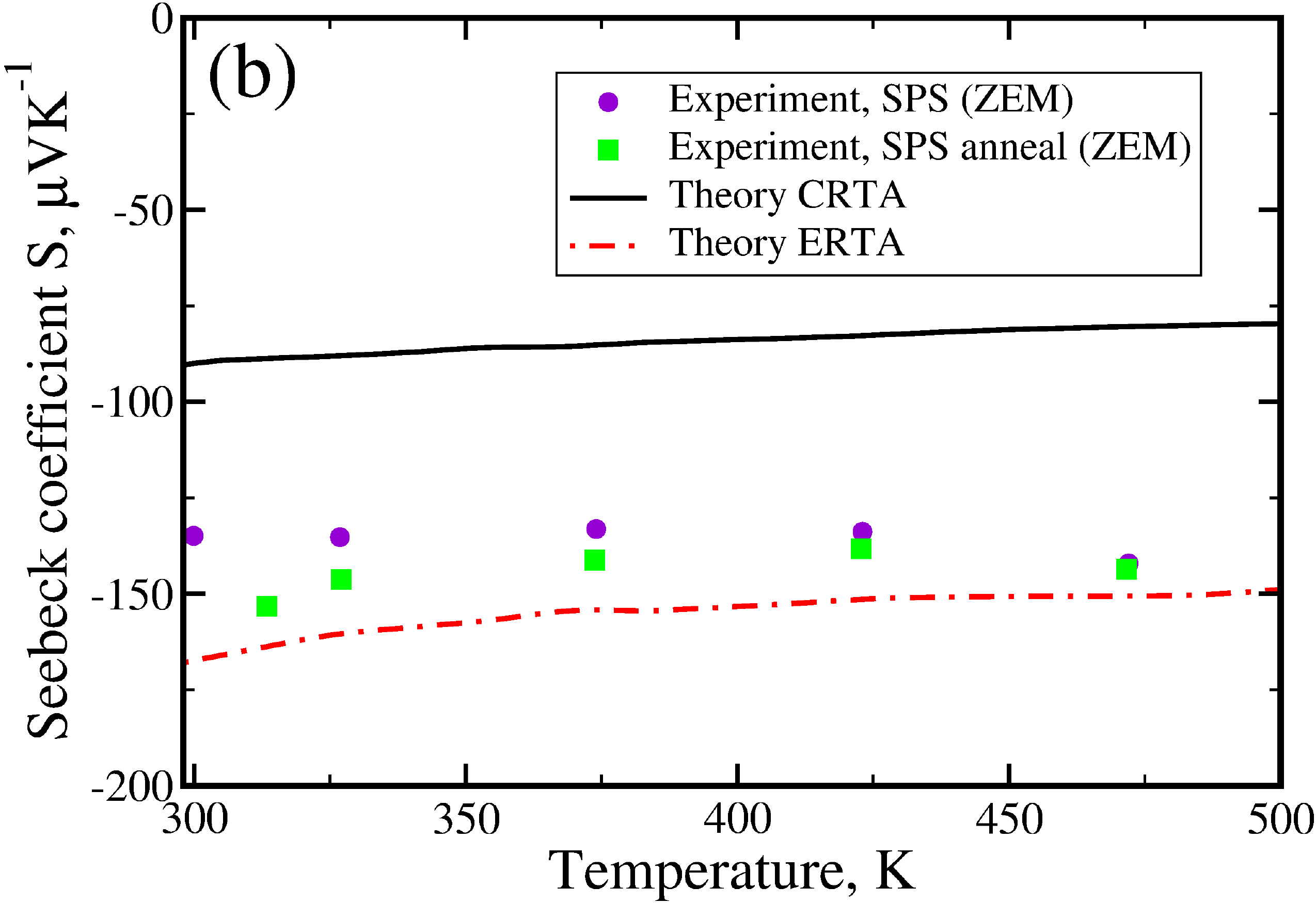}}
\caption{\textbf{Panel (a)}: Temperature variation of the electrical conductivity $\sigma$ measured in the experiment in $n$-type samples before (violet circles) and after (red and green squares) 
annealing. The fitting curves are shown by dashed blue and solid black lines respectively. \textbf{Panel (b)}: Temperature variation of the Seebeck coefficient $S$ for $n$-type samples
measured in experiment (violet circles and green squares). The theoretical Seebeck coefficients calculated in the CRTA and in the ERTA 
are shown by black solid and red dashed dotted lines respectively. We have used ZEM and PPMS systems for the measurements.}
\label{fig:elect_transport_temp}
\end{figure*}

In Fig.~\ref{fig:bandstructure} (a), we compare the electronic band structures calculated using GGA-PBE~\cite{Perdew:1996} (black curves) and hybrid-HSE06~\cite{Heyd:2003} (red curves)
exchange-correlation functionals and summarize the theoretical and experimental band edges, E, and spin-orbit splittings, $\Delta$, at $\Gamma$, L and X high symmetry points in Table~\ref{tab:bandstructure}.
First, we note that the HSE06 calculation predicts the correct level ordering $\Gamma_7$, $\Gamma_6$, $\Gamma_8$~\cite{Svane:2011,Nicklas:2011} that is consistent with experiment~\cite{Orlowski:2000}
in contrast to the GGA-PBE calculation where the $\Gamma_6$ and $\Gamma_7$ bands are reversed. Second, the band energies obtained with the hybrid functional are in excellent agreement with 
experiment. For instance, the inverted band gap $E_g = E_{\Gamma_6} - E_{\Gamma_8} = -0.27$ eV and spin-orbit splitting $\Delta_0 = E_{\Gamma_8} - E_{\Gamma_7} = 0.89$ eV at $\Gamma$ differ from their
experimental values only by 0.02 eV. Third, the effective mass of the lowest conduction band is significantly reduced from $m_e$ = 0.18 $m_0$ in GGA-PBE to $m_e$ = 0.04 $m_0$ in HSE06 in the [100] direction, 
whereas the effective mass of the top valence bands remains essentially unchanged $m_h$ = 0.29 $m_0$ in GGA-PBE to $m_h$ = 0.33 $m_0$ in HSE06 . Thus, HgTe is a material with a very high electron-hole
effective mass ratio.

Finally, the electronic properties of HgTe near the Fermi level are defined by the region of the Brillouin zone close to the $\Gamma$ point, where the bands have a low degeneracy.
This low degeneracy in combination with a small electron effective mass in HSE06 calculation results in a small density of states of conduction bands.
The asymmetry between the conduction and valence bands is clearly seen in both, the density of states $g(E)$ and the differential conductivity $\sigma_{xx}(E)$, as can be seen in Fig.~\ref{fig:bandstructure} (b) and (c) respectively.

In Fig.~\ref{fig:elect_transport_300K} (a), we show the Seebeck coefficient as a function of doping concentration for $p$- and $n$-types of doping at T = 290 K calculated 
using the constant relaxation time approximation. Our results with the GGA-PBE functional agree well with the previous calculation of Chen \textit{et al.}~\cite{Chen:2008} done with 
the same exchange-correlation potential. As it is expected from the band structure calculations, one can see a noticeable change in the magnitude of the Seebeck coefficient due to 
the increase of the electron-hole effective mass ratio in HSE06 calculation. For instance, the maximum of the Seebeck coefficient is increased from 142 $\mu$V/K to 202 $\mu$V/K and 
is slightly shifted towards the lower doping concentrations from $2\cdot 10^{19}$ cm$^{-3}$ to $9\cdot10^{18}$ cm$^{-3}$. In intrinsic and low doped samples (up to $10^{17}$ cm$^{-3}$),
the Seebeck coefficient remains constant but also has a sufficiently higher magnitude of -81 $\mu$V/K with HSE06 instead of $-31$ $\mu$V/K with GGA-PBE. Our HSE06 result is in good
agreement with experimental result -91 $\mu$V/K (blue circle) reported by Whitsett \textit{et al}~\cite{Whitsett:1972} for p-type sample. However, our measurements in n-type HgTe 
sample with $n = 3.5\cdot10^{17}$ cm$^{-3}$ doping concentration show much larger values of the Seebeck coefficient of -136 $\mu$V/K (black square).

The constant relaxation time theory, does not allow to compute the electrical conductivity but only its ratio to the unknown relaxation time $\frac{\sigma}{\tau}$. As can be seen in
Fig.~\ref{fig:elect_transport_300K} (b), this ratio varies slowly at low doping concentrations and grows rapidly at high doping concentrations. However, one would expect a different 
behavior for the electrical conductivity at least in the high doping concentration region where a strong charged carrier scattering limits the mobilities. Thus, to further investigate
the behavior of the electrical conductivity and the Seebeck coefficient, we introduce the phenomenological scattering rates and fit them to reproduce our experimental
electrical conductivity data in $n$-type sample.

Fig.~\ref{fig:elect_transport_temp} (a) show our experimental data obtained using the four-terminal probe method~\cite{Smits:1958} in the samples prepared using the spark plasma sintering (SPS) 
technique. Two sets of measurements before (violet circles) and after (red and green squares) annealing have been performed. As expected,
annealing improves the electrical conductivity~\cite{Okazaki:1975,Whitsett:1972} which reaches its maximum value of $\sigma = 1036$ ($\Omega$ cm)$^{-1}$ at 
T = 350 K and then starts monotonically decreasing at higher temperatures. We notice that our results are much lower than the electrical conductivity $\sigma = 1700$ $\Omega^{-1}$cm$^{-1}$
measured in the intrinsic samples at T = 300 K~\cite{Dziuba:1964,Whitsett:1972}. These intrinsic samples were prepared by multiple annealing of the originally $p$-type samples in the presence
of Hg gas~\cite{Dziuba:1964,Whitsett:1972}. However, in this work we do not follow this procedure due to the extreme toxicity of mercury. 

We fit the measured electrical conductivity using \textit{ab initio} data for the differential conductivity $\sigma_{xx}(E)$ and the density of states $g(E)$ obtained with the
hybrid-HSE06 functional and phenomenological energy dependent scattering rates accounting for the acoustic deformation potential, polar optical and ionized impurity scattering rates.~\cite{Lundstrom:2000} Details of the considered scattering rates are described in Supplementary information.
We then recalculate the Seebeck coefficient using the obtained scattering rates and find that its magnitude is increased about 2 times with respect to the constant relaxation time
approximation (CRTA). The energy dependent relaxation time approximation (ERTA) results in Seebeck coefficient values that are closer to the experimentally measured ones. Therefore 
we conclude that the difference between the CRTA calculations (Fig. 2a) and experimental values is a result of the energy dependence of the scattering rates. Although the Seebeck 
coefficient is not as sensitive as the conductivity to the relaxation times, this example demonstrates that CRTA results could be misleading even in calculation of the Seebeck 
coefficient. 

 The temperature variation of the Seebeck coefficient calculated in the CRTA (black solid lines), the ERTA (red dashed dotted line) and measured in experiment are shown in
Fig.~\ref{fig:elect_transport_temp} (b). As one can see, both the theoretical and experimental Seebeck coefficients remain almost temperature independent in the studied temperature range between 300
and 500 K. 

Our study reveals that for the accurate description of the electrical transport properties of HgTe, one needs to accurately reproduce the electron-hole effective mass ratio that 
can not be achieved using standard LDA or GGA exchange-correlation functionals. Moreover, we find that the inclusion of energy dependent scattering rates changes the magnitude of the Seebeck 
coefficient drastically. The latter has been unexpected since, according to the common believe~\cite{Madsen:2006}, the CRTA reproduces well the behavior of the diffusion part of the Seebeck 
coefficient. The magnitude of the Seebeck coefficient of HgTe is an order of magnitude higher than the one in typical metals and close to the typical values of
narrow-gap semiconductors. That is explained by the the low effective mass and low degeneracy of the conduction band near the Fermi level. We then conclude that the presence of a bandgap is not essential for obtaining large Seebeck coefficient values.   

\subsection{Thermal transport}

\begin{figure}[t]
\centering
\includegraphics[width=0.8\linewidth]{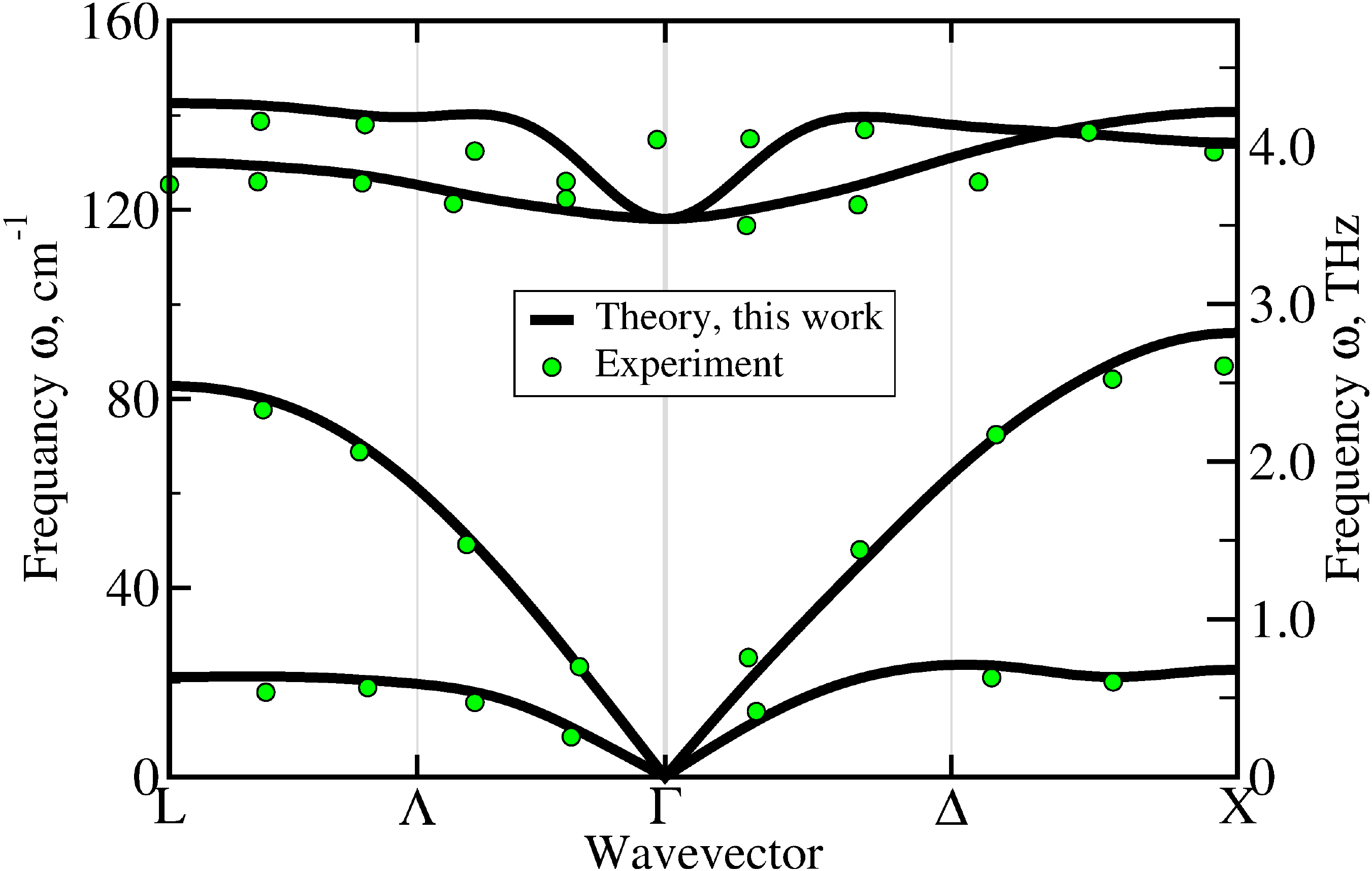}
\caption{Theoretical phonon dispersion calculated using DFPT in this work (black curves) compared to the inelastic neutron scattering data (green circles)~\cite{Kepa:1980,Kepa:1982}.}
\label{fig:ph_disp}
\end{figure}

Now, we turn our attention to the thermal transport properties of HgTe. First, we investigate the lattice dynamics by calculating the phonon spectrum along the high symmetry directions. The 
phonon dispersion is shown in Fig.~\ref{fig:ph_disp} and is in an excellent agreement with previous theoretical results~\cite{Cardona:2009,Ouyang:2015,Ouyang:2015a} as well as with 
available data from the inelastic neutron scattering experiments~\cite{Kepa:1980,Kepa:1982} (green circles). In our calculations we do not take into account the non-analytical 
correction to split the optical phonons at $\Gamma$ point. However, this correction should not strongly affect the thermal conductivity since the contribution is usually small due to the
low group velocities of optical phonons. Our theoretical frequencies for optical phonons $\omega_{O}(\Gamma) = 118$ cm$^{-1}$ agree well with the Raman spectroscopy data 
for the transverse optical phonons $\omega_{TO}(\Gamma) = 116$ cm$^{-1}$~\cite{Mooradian:1971}.    

To further validate the vibration spectrum, we calculated the elastic constants $C_{ij}$. As shown in Table~\ref{tab:elastic}, the difference between our theoretical results 
and experiment does not exceed 4\%. Then, we compare the sound velocities in [100] direction obtained from the elastic constants, from the slopes of acoustic branches near the $\Gamma$ point and experimental 
data in Table~\ref{tab:svelocity}. The largest differences with the experiment, 7.1\% and 2.5\% for the transverse (TA) and longitudinal (LA) sound velocities respectively, are found for the 
evaluation of sound velocities from the slopes of acoustic phonons.  

\begin{table}
\centering
\begin{tabular}{clll}
\hline
            & $C_{11}$, GPa & $C_{12}$, GPa & $C_{44}$, GPa \\
\hline
 Present    &   57.3   &   41.0   &   22.0   \\
 Experiment &   59.7~\cite{Madelung:1982}   &   41.5~\cite{Madelung:1982}   &   22.6~\cite{Madelung:1982}  \\
 \multirow{2}{*}{Other}      &   56.3~\cite{Rajput:1996}   &   37.9~\cite{Rajput:1996}   &   21.2~\cite{Rajput:1996}  \\
            &   67.4~\cite{Tan:2010}   &   45.7~\cite{Tan:2010}   &   30.0~\cite{Tan:2010}  \\
\hline
\end{tabular}
\caption{\label{tab:elastic}Elastic constants $C_{ij}$ (GPa) calculated in the present work and compared with other theoretical calculations~\cite{Rajput:1996,Tan:2010} and experiment~\cite{Madelung:1982}.}
\end{table}
\begin{table}
\centering
\begin{tabular}{cll}
\hline
              & $v_{L}$, m/s & $v_{T}$, m/s \\
\hline
 Elastic constant     &   2655   &   1645 \\
 Slope                &   2747   &   1504 \\
 Experiment           &   2680   &   1620 \\
\hline
\end{tabular}
\caption{\label{tab:svelocity}The longitudinal $v_{L}$  and transverse $v_{T}$ sound velocities (m/s) in [100] direction calculated in the present work from the elastic constants, 
slopes of acoustic phonons and experiment.}
\end{table}

\begin{figure*}[t]
\hfill
\subfigure{\includegraphics[width=0.49\linewidth]{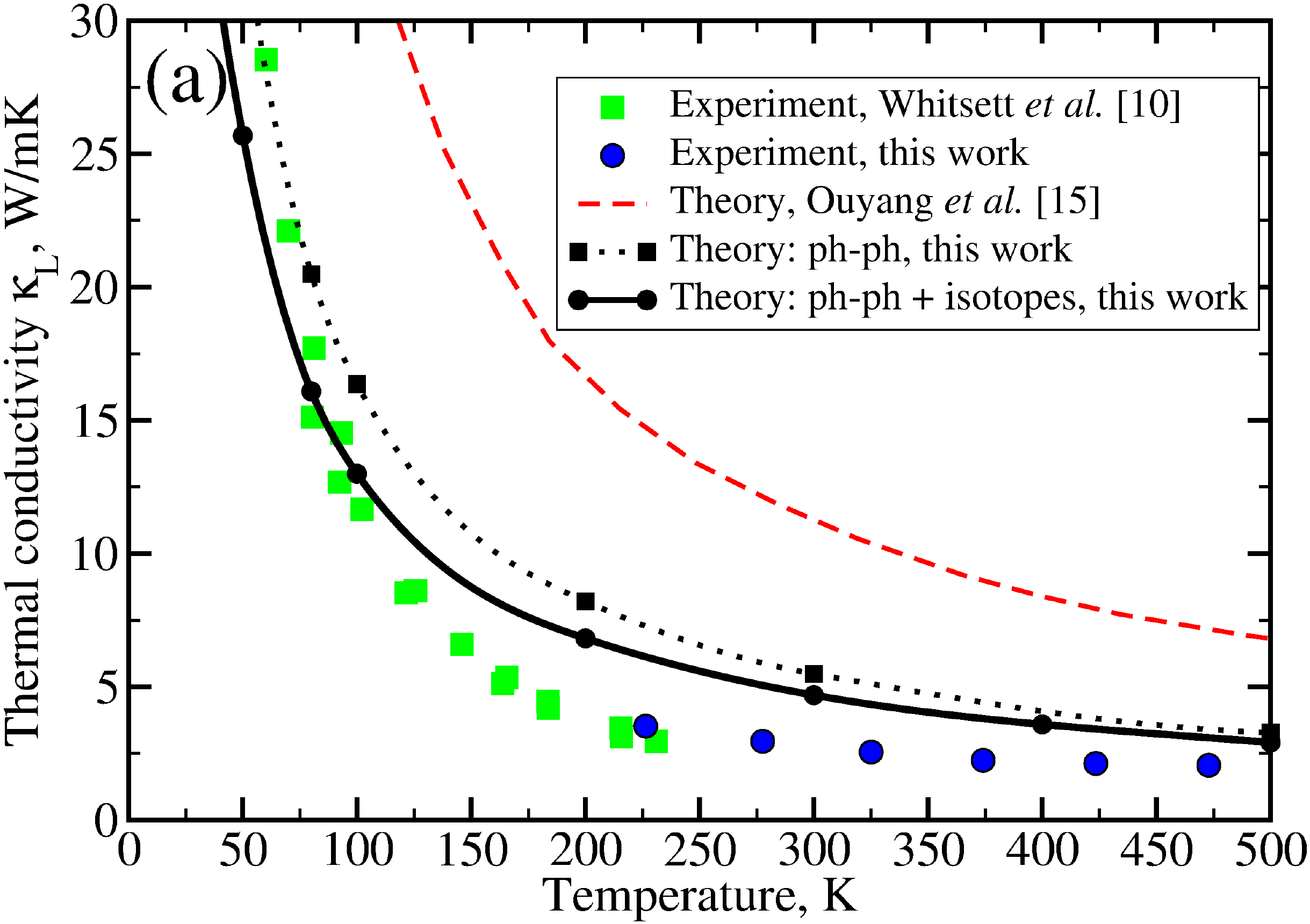}}
\hfill
\subfigure{\includegraphics[width=0.49\linewidth]{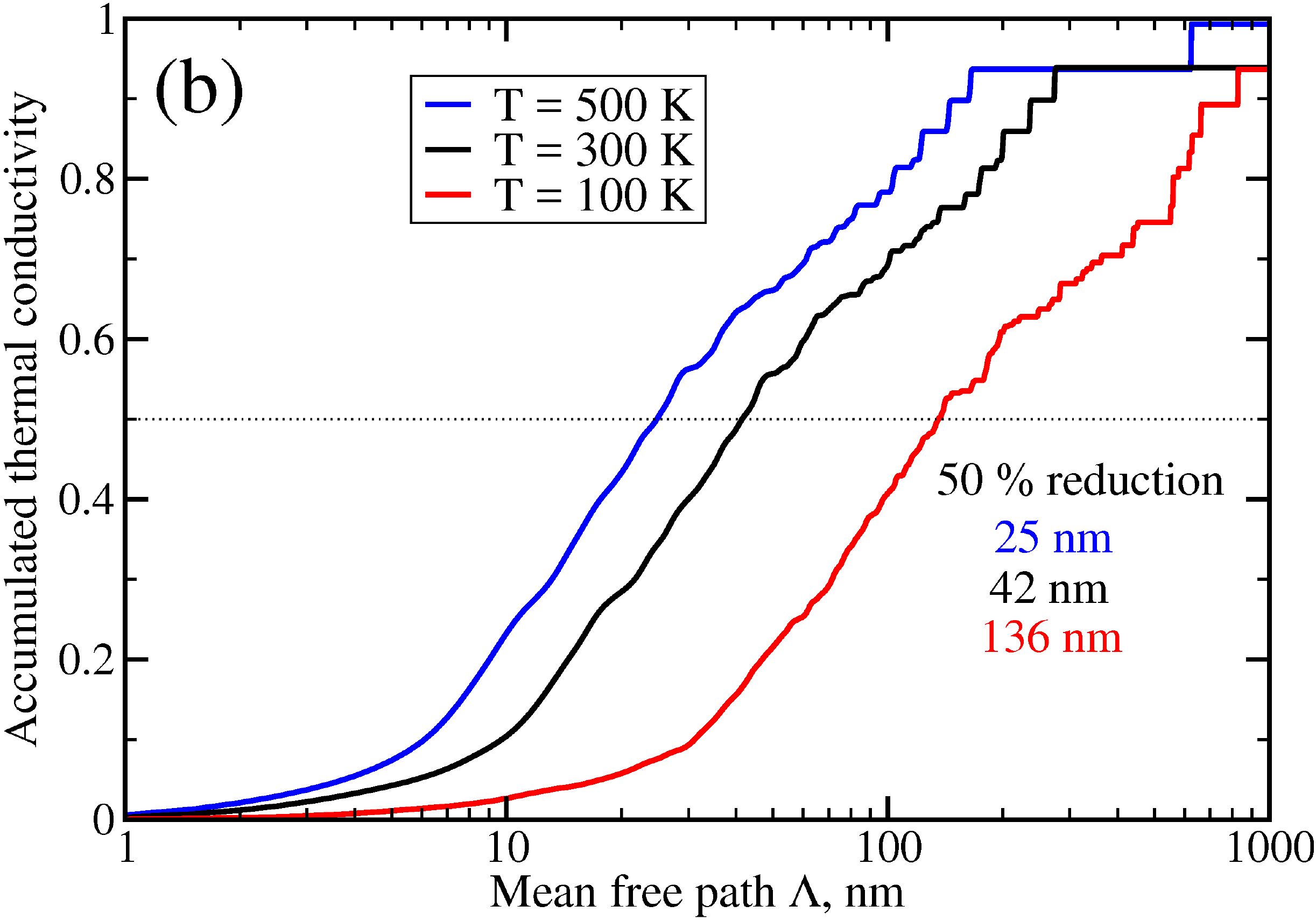}}
\caption{
\textbf{Panal (a):} Temperature dependence of the thermal conductivity calculated with account for anharmonic three-phonon processes only (black dashed line ) and with addition of isotopic 
disorder scattering (black solid line); green squares - experimental data from Whitsett \textit{et al}~\cite{Whitsett:1972}; blue circles - our experimental data; dashed red curve - previous computational result from Refs~\cite{Ouyang:2015,Ouyang:2015a}.
\textbf{Panal (b):} Accumulated thermal conductivity as a function of phonon mean free path $\Lambda$ at T = 100 K (blue curve), 300K (black curve) and 500K (red curve). Horizontal dotted line denotes 50 \% thermal 
conductivity reduction.
}
\label{fig:thermalcond}
\end{figure*}

\begin{figure*}[t]
\hfill
\subfigure{\includegraphics[width=0.49\linewidth]{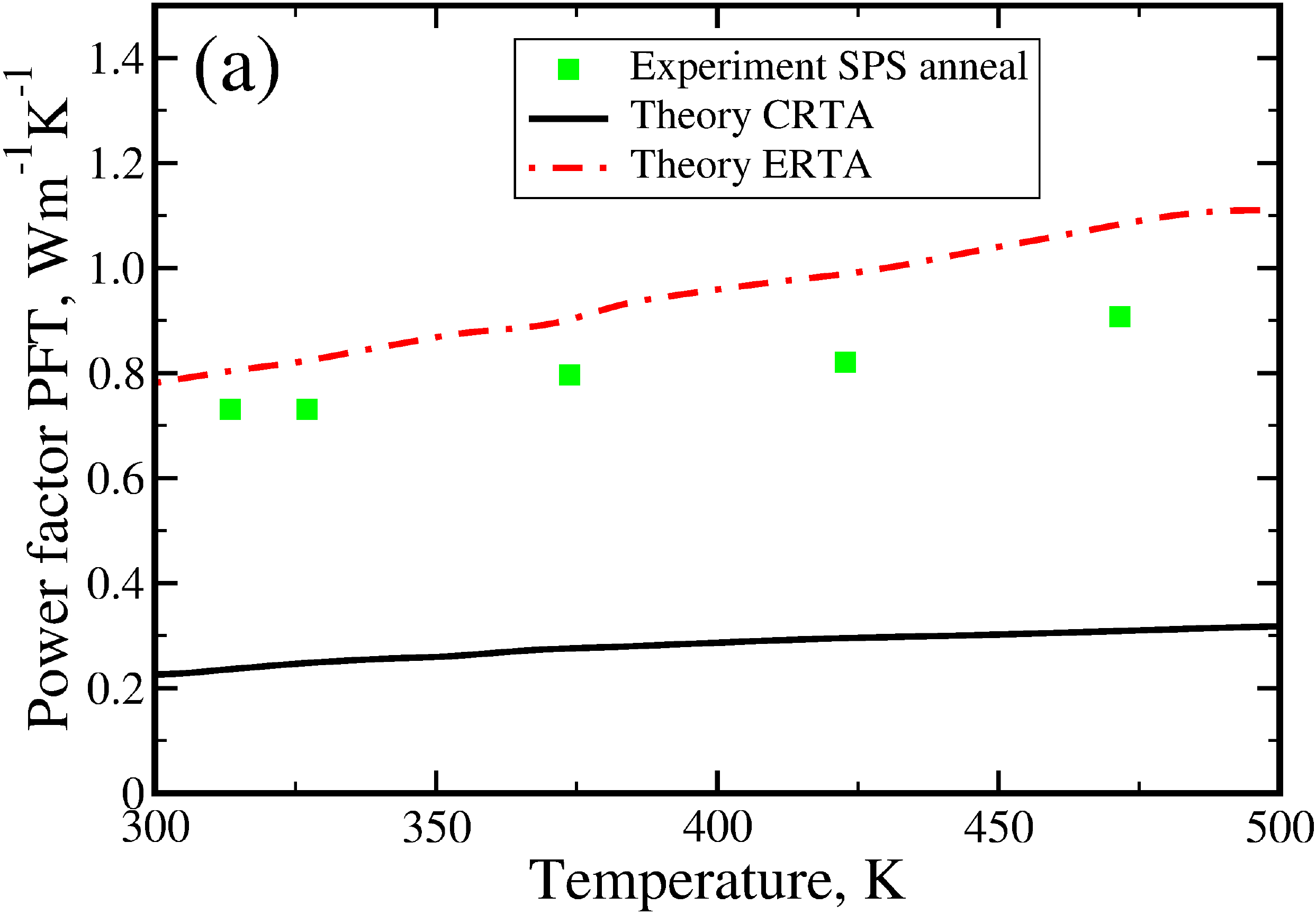}}
\hfill
\subfigure{\includegraphics[width=0.49\linewidth]{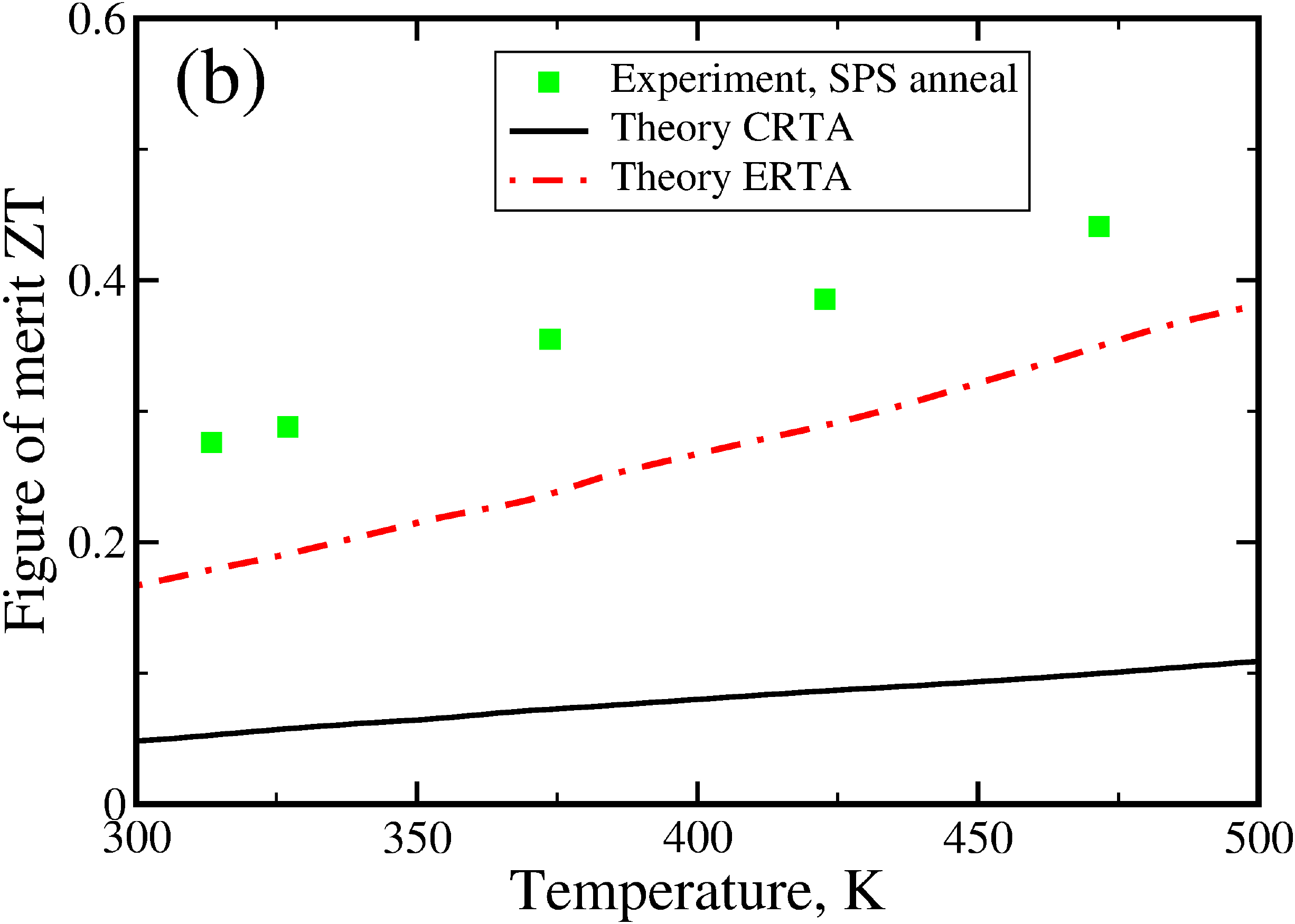}}
\caption{Temperature variation of the power factor $PFT = S^2\sigma T$ (panel a) and thermoelectric figure of merit $ZT$ (panel b) measured in the experiment (green squares) and calculated 
in the CRTA (black solid line) and in the ERTA (red dashed dotted line).}
\label{fig:thermoelectric}
\end{figure*}

Figure~\ref{fig:thermalcond} summarizes the theoretical and experimental thermal conductivity obtained in this work as well as those reported by other groups. We perform the lattice 
thermal conductivity calculations by \textit{exactly} solving the Boltzmann Transport Equation (BTE). First, we include only the intrinsic three-phonon anharmonic scattering 
(dotted black curve). We obtain the lattice thermal conductivity that is much lower than the previous \textit{ab initio} calculations (red dashed curve)~\cite{Ouyang:2015}.
For instance, we get $\kappa_L = 5.48 $ W/mK instead of $\kappa_L = 10.46$ W/mK in Ref.~\cite{Ouyang:2015}. Our theoretical values are still higher than ones measured in experiment 
$\kappa_L = 2.9$ W/mK (this work) or $\kappa_L = 2.14$ W/mK (Ref.~\cite{Whitsett:1972}). This discrepancy can not be attributed to the extrinsic sources of scattering such as the impurity 
scattering since the experimental data for the $p$-type samples with doping concentration between $10^{16}-10^{18}$ cm$^{-3}$ show essentially the same thermal conductivity~\cite{Whitsett:1972}.
The addition of isotopic disorder scattering significantly decreases the thermal conductivity mainly at low temperatures (black solid curve) whereas at high temperatures the isotopic 
scattering plays a minor role. At room temperature we get $\kappa_L = 4.68$ W/mK that is still higher than experimental values. 

While we capture the low temperature trend, we attribute the disagreement between experiment and theory at higher temperatures to some intrinsic scattering mechanism which has not been taken into account in our calculations. We assume that four-phonon anharmonic processes 
or higher order three phonons are important because of the deviations of $\kappa(T)$ from the 1/T behavior. Thus, the lattice thermal conductivity of HgTe should be subject to further investigation.

In Fig.~\ref{fig:thermalcond} b, we analyze the accumulated lattice thermal conductivity $\kappa_L(\Lambda)$ as a function of phonon mean free path $\Lambda$ (see supplementary 
material for details) at three different temperatures T = 100 K (blue curve), 300 K (black curve), 500 K (red curve). As one can see, the thermal conductivity is mainly cumulated below 1 micron and the mean free paths become shorter when temperature is increased.
The accumulated function can be used to predict the effective size $L$ of a nanostrucure necessary to reduce the thermal conductivity and, thus, increase the thermoelectric performance of 
a material. Indeed, phonons with mean free paths larger than $L$ are scattered by sample boundaries and their contribution to the thermal conductivity is suppressed.
The horizontal dotted line denotes a 50\% reduction of thermal conductivity. It is found to be $L = 136$ nm at T = 100 K, $L = 42$ nm at T = 300 K and $L = 25$ nm at T = 500 K.

\subsection{Thermoelectric performance}

Finally, we evaluate the overall thermoelectric power factor $PFT = S^2 \sigma T$ based on our experimental and theoretical data in Fig.~\ref{fig:thermoelectric} (a). As one can see,
HgTe possess a high power factor which grows with temperature linearly from 0.8 W m$^{-1}$ K$^{-1}$ at T = 310 K to 0.9 W m$^{-1}$ K$^{-1}$ at T = 475 K. Our theoretical values obtained 
in the ERTA slightly overestimate the experimental power factor but show the same temperature dependence reaching 1.0 W m$^{-1}$ K$^{-1}$ at T = 500 K. The CRTA underestimates the magnitude of the 
Seebeck coefficient and results in a low power factor around 0.2 W m$^{-1}$ K$^{-1}$.The figure of merit also increases linearly since thermal conductivity is relatively unchanged in this temperature range. 

While ZT values reported here are small. We would like to emphasize that this is not an optimized sample. One can increase the ZT values by many different techniques. For example, further increase in the electrical conductivity (a factor of two) is expected after 
annealing in Hg gas with relatively unchanged Seebeck coefficient and thermal conductivity values~\cite{Dziuba:1964,Whitsett:1972}. As mentioned earlier we avoid this process due to both toxicity of Hg gas and the 
fact that optimization of the thermoelectric properties of HgTe is not the subject of this work. 
One can also implement nanostructuring to further reduce the thermal conductivity, a technique that is routinely performed to optimize the thermoelectric figure of merit. Similarly,
slight doping (tunning of the chemical potential) and slight alloying could be used to further optimize the performance of semimetallic HgTe. For example, alloying with cadmium could 
lower the thermal conductivity and still preserves the semimetallic nature of the HgTe for small molar fractions of cadmium ($x<0.1$).    

\section{Methods}

\subsection{Theoretical methods}

Our theoretical calculations are based on density functional theory (DFT). For the electrical transport calculations, we use Vienna Ab-initio Simulation Package 
(VASP)~\cite{Kresse:1996,Kresse:1996a} combined with Boltzmann Transport Theory as implemented in Boltztrap code~\cite{Madsen:2006}. We use pseudopotentials based
on the projector augmented wave method~\cite{Blochl:1994} from VASP library with the generalized gradient approximation by Perdew, Burke and Ernzehof (GGA-PBE)~\cite{Perdew:1996}
and with a hybrid Heyd-Scuseria-Ernzehof (HSE06)~\cite{Heyd:2003} exchange-correlation functionals. A plane wave kinetic cut-off of $E_{cut} = 350$ eV and $\Gamma$-centered 
k-point mesh of 8x8x8 were found to be enough to converge the total energy up to 5 meV~\cite{Nicklas:2011,Nicklas:2013}. We use a tetrahedron method for the Brillouin zone integration
and the experimental lattice parameter a = 6.460 $\AA$ in both calculations. In our calculations, we take into account the spin-orbit coupling which is important to accurately reproduce
the electronic band structure of HgTe. To ensure the convergence of transport integrals in Boltztrap, we use 20 times denser interpolated grid than we do in our ab initio calculations.

For the thermal transport calculations, we use Quantum Espresso~\cite{Giannozzi:2009} package combined with D3Q code to calculate third-order anharmonic force constants using "2n+1" theorem~\cite{Paulatto:2013} 
and to solve the Boltzmann Transport equation for phonons variationally~\cite{Fugallo:2013}. We use the norm-conserving pseudopotentials with the exchange-correlation part treated in 
the local density approximation by Perdew and Zunger (LDA-PZ)~\cite{Perdew:1981}. We use a cut-off energy of $E^{cut}$ = 1360 eV (100 Ry), 8x8x8 k-points mesh to sample the Brillouin zone
with Methfessel-Paxton smearing of $\sigma$ = 0.068 eV (0.005 Ry). The equilibrium lattice parameter is found to be $6.431$ $\AA$. Spin-orbit coupling is not included in the 
calculations since it has a weak effect on vibrational properties of HgTe as has been pointed out by M. Cardona \textit{et al.}~\cite{Cardona:2009}. Phonon frequencies and group
velocities are calculated using the density functional perturbation theory (DFPT)~\cite{Baroni:2001} on a 8x8x8 $\textbf{q}$-point grid centered at $\Gamma$. The third-order anharmonic
constants are calculated on a 4x4x4 $\textbf{q}$-point grid in the Brillouin zone that amounts to 42 irreducible triplets. Both phonon harmonic and anharmonic constants are then
interpolated on a dense 24x24x24 $\textbf{q}$-point grid necessary to converge the thermal conductivity calculations. 

The detailed information about the charged carrier scattering rates obtained from the electrical conductivity fit and about the isotopic disorder scattering rates used in the lattice thermal conductivity calculation
is reported in the supplementary material.

\subsection{Experimental methods}

A 99.99\% purity of HgTe ingot was purchased from 1717 CheMall Corporation for HgTe sample preparation and the density of the ingot was 7.82 $\pm$ 0.04 g/cm$^{3}$ obtained by Archimedes$'$
 principle. We crashed the ingot and milled it with a mortar and pestle for about 10 minutes to obtain fine powders. Later, they were consolidated into a 0.5$''$-diameter compact disk by 
using spark plasma sintering (SPS) method at 783K, 50MPa for 15 minutes. After SPS process, the density of the HgTe disk is increased to 7.98 $\pm$ 0.17 g/cm$^{3}$, which is
quite close to the theoretical fully-dense value of the HgTe density 8.13 g/cm$^{3}$. For annealing preparation, the compact HgTe sample was sealed in an evacuated capsule, and it was situated in
the middle of a furnace at 523 K for 5 days.

\begin{figure}[p]
\centering
\includegraphics[width=0.95\linewidth]{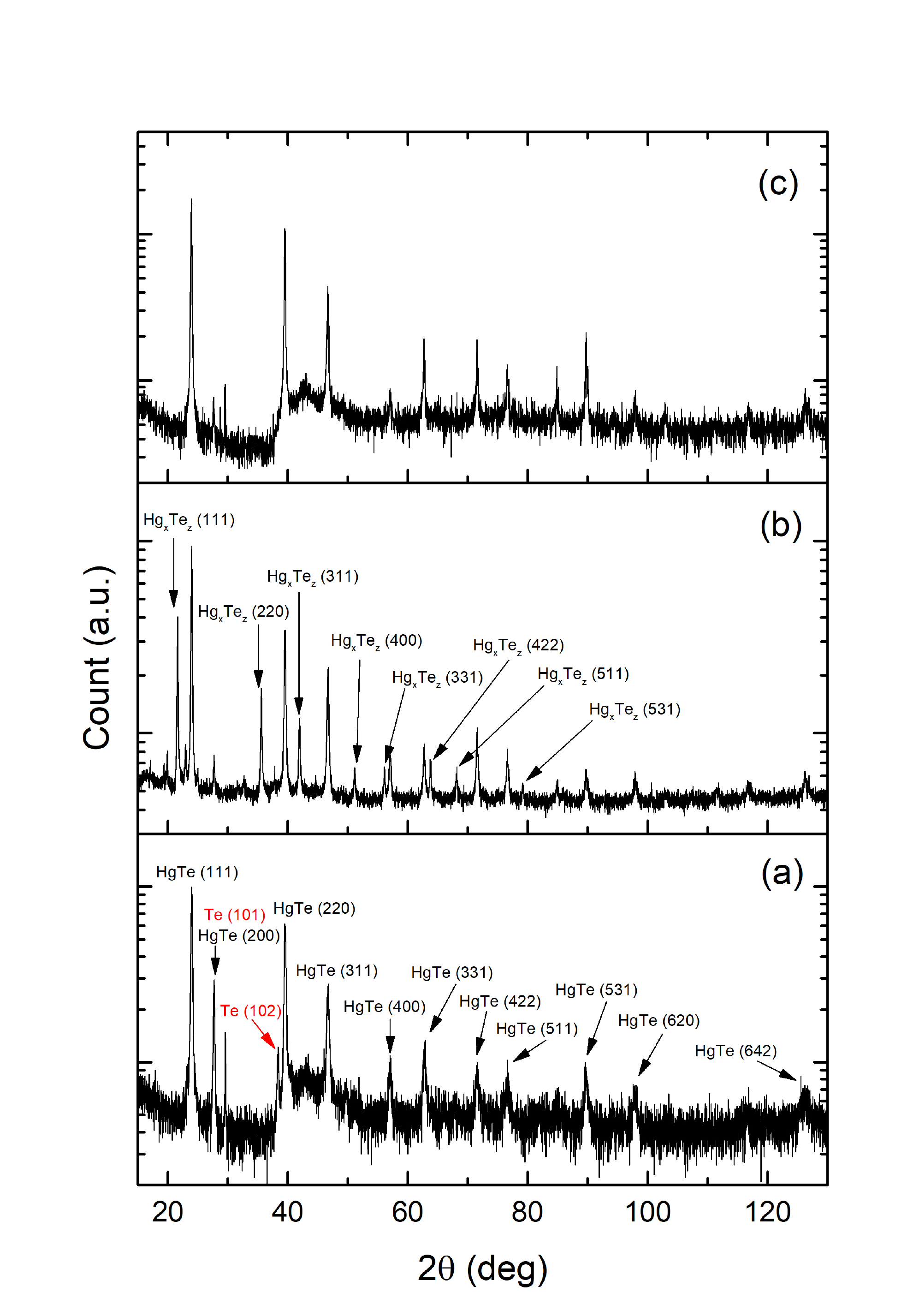}
\caption{The comparison of the x-ray diffraction (XRD) results between (a) ingot, (b) SPS, and (c) SPS-annealing samples. The excess Te peaks in the ingot samples are highlighted by red color.} 
\label{fig:xray}
\end{figure}

For the ingot samples, we used the machine to cut it into a rectangular-bar-shaped sample with the dimension of 2x4x8 mm$^{3}$. For the SPS samples, due to their fragality, we hand-polished the disk into a 
rectangular-shaped bar of the same size as the ingot one instead of cutting them in the machine. The four-probe electrical conductivity and Seebeck coefficient measurements were performed in 
the helium atmosphere with a ZEM-3 equipment from Ulvac Tech., Inc. The Hall coefficient measurements were conducted in Quantum Design Versa-Lab. The thermal diffusivity 
experiments were carried out with a LFA 467 HyperFlash equipment from NETZSCH. The measured thermal diffusivity was then multiplied by the theoretical heat capacity~\cite{Glazov:1996}
$C_{p}(T) = C_{V}(T) + 1.01\cdot10^{-2}T$ where $C_V(T)$ was obtained from the Debye model.

X-ray diffraction data are shown in fig.~\ref{fig:xray}.  Figure.~\ref{fig:xray}(a) shows the original ingot contains single-phase HgTe with excess tellurium. After the SPS process, 
in additional to the original HgTe phase, a new crystal phase, Hg$_x$Te$_z$, emerges (see panel (b)), and the excess Te peaks, i.e., Te (101) and Te (102), disappear. Panel (c) shows
that the SPS-annealing sample is a single-phase HgTe crystal without the excess of Te. Note that the phase of Hg$_{x}$Te$_{z}$ vanishes after annealing.

Additional information about the Hall coefficient measurements is provided with the supplementary material.

\section{Conclusions}

In conclusion, we have investigated, both experimentally and theoretically, the electrical and thermal transport properties of HgTe at high-temperatures between 300 and 500 K.
We have found that HgTe is a good thermoelectric material in a low pressure semi-metallic zinc-blende phase as it has a high Seebeck coefficient and a low thermal conductivity.
To explain the experimental data for the Seebeck coefficient, we accurately reproduce the electron-hole effective mass ratio by performing \textit{ab initio} calculations with
the hybrid-HSE06 exchange-correlation functional and take into account the phenomenological scattering rates extracted from a fit to electrical conductivity. Finally, we perform 
the lattice thermal conductivity calculations by \textit{exactly} solving the Boltzmann Transport Equation (BTE). We include three-phonon anharmonic scattering and isotopic disorder 
scattering processes. We attribute the disagreement between experiment and theory to some intrinsic scattering mechanism which has not been taken into account in our calculations. 
Our work demonstrates that large thermoelectric power factors could be achieved even in the absence of an energy bandgap. 

\section*{Acknowledgements}

M.M., H.L and M.Z acknowledges support from Air Force Young Investigator Award (Grant FA9550-14-1-0316). M.Z. acknowledges support from National Space Grant College and Fellowship 
Program (SPACE Grant) Training Grant 2015-2018, grant number NNX15A120H. We acknowledge the SEAS for the computational time on Rivanna HPC cluster.

\section*{Conflict of interest}
There are no conflicts to declare.


\clearpage

\setcounter{figure}{0}
\renewcommand{\theequation}{S.\arabic{equation}}
\renewcommand{\thefigure}{S\arabic{figure}}

\makeatletter
\renewcommand*{\@biblabel}[1]{\hfill[S#1]}
\makeatother

\section*{\centering Supplementary material \\ for \\ "Semi-metals as potential thermoelectric materials: case of HgTe"}

In this supplementary material, we provide the supporting information about the Hall coefficient measurements, the electrical conductivity fitting, the lattice thermal conductivity 
calculations and measurements.

\subsection{Hall coefficient measurements.}
\begin{figure}[p]
\centering
\includegraphics[width=1.0\linewidth]{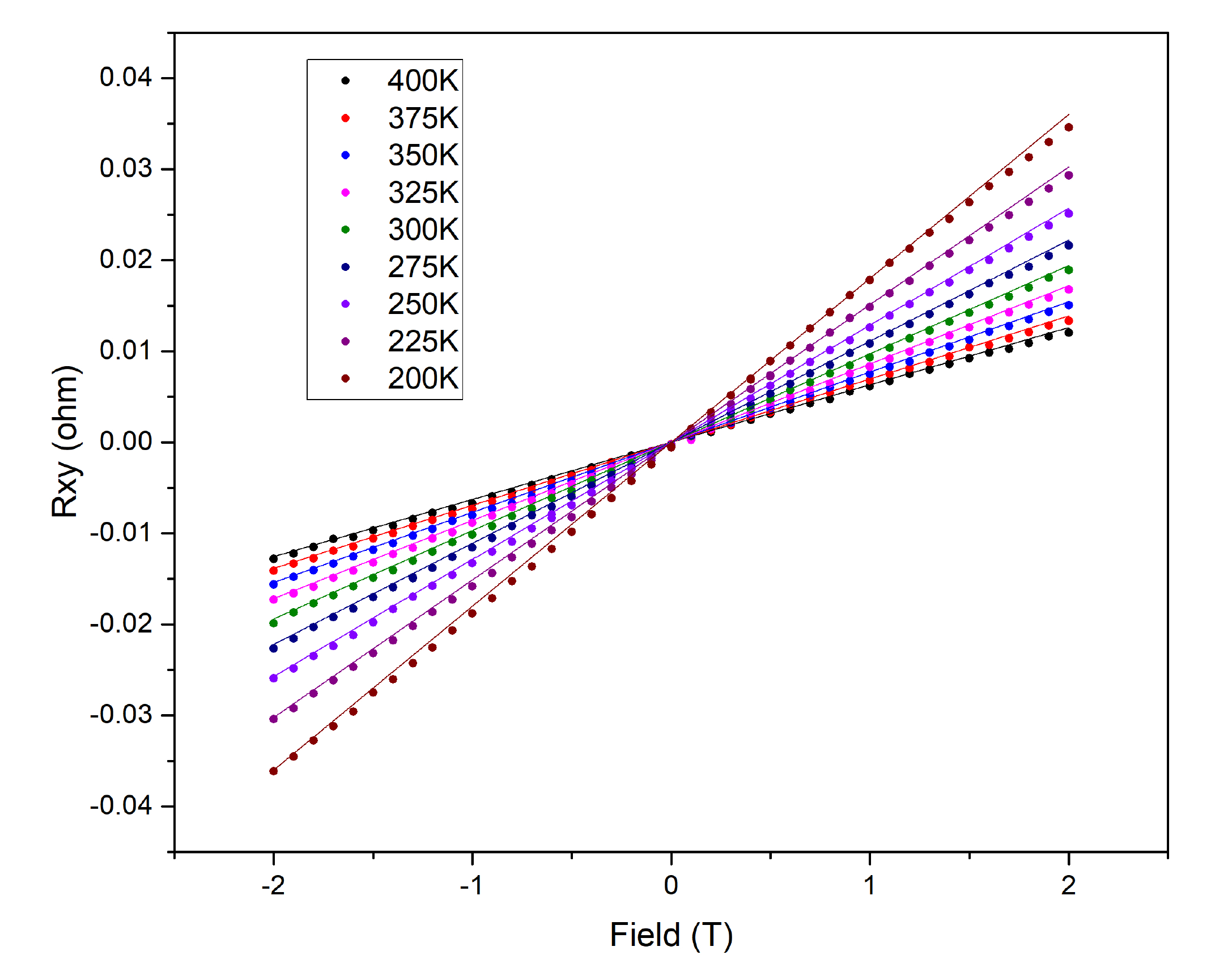}
\caption{The Hall effect resistance $R_{xy}$ measured as a function of magnetic field $B$ at different temperatures.}
\label{fig:Hall}
\end{figure}

In Fig.~\ref{fig:Hall}, we show the experimental data for the Hall effect resistance $R_{xy}$ as a function of an applied external magnetic field $B$ in the temperature range 200 K $\leq$ T 
$\leq$ 400 K. The Hall coefficient $R_H$ can be extracted from the slope of the $R_{xy}(B)$ curve as
\begin{equation}
 R_{H}(T) = \frac{R_{xy}(T)}{B}l
\end{equation}
where $l$ is the sample thickness. The net carrier concentration and electron mobilities can be found from the Hall coefficient data as
\begin{equation}
 n (T) = \frac{1}{eR_H (T)} 
\end{equation}

\begin{equation}
 \mu_e (T) = \frac{R_{H}(T)}{R l}
\end{equation}
and are shown in Fig.~\ref{fig:concentration} and Fig.~\ref{fig:mobility} respectively. Here $e$ is an elementary charge, $R$ is the resistance of the sample without magnetic field. 

\subsection{Electrical conductivity fitting.}
The electrical conductivity can be found using the following expression
\begin{equation}
\sigma(T,\mu) = \frac{1}{V_{cell}} \int \sigma(E) \left[-\frac{\partial f_{\mu} (T,E) }{\partial E}\right]dE
\end{equation}
where $V_{cell}$ is a unit cell volume, $E$ is energy, $\mu$ is the chemical potential, $f_{\mu}$ is the Fermi-Dirac distribution function and $\sigma(E)$ is the differential
conductivity
\begin{equation}
 \sigma(E) = e^2 \tau(E)g(E)v_g^2(E)
\end{equation}
where $g(E)$ is the density of states, $v_g(E)$ is a group velocity and $\tau(E) = 1/\Gamma(E)$ is the total relaxation time that is inversely proportional
to the total scattering rate $\Gamma(E)$. In our calculations we use $g(E)$ and $v_g$ obtained with the HSE06 exchange-correlation functional.
In the constant relaxation time approximation (CRTA), one assumes that $\tau(E)$ is constant and energy independent.

In this work, we consider the energy dependent scattering rates. We consider 3 types of carrier scattering including acoustic deformation potential scattering $\Gamma^{ac} (E)$, 
ionized impurity scattering $\Gamma^{imp} (E)$ and polar optical scattering $\Gamma^{pop} (E)$~\cite{Lundstrom:2000}. As follows from the Matthiessen's rule, the total scattering rate
$\Gamma(E)$ is a sum of all three contributions. Overall, we have 4 fitting parameters $A_1$, $A_2$, $A_3$ and phonon energy $\hbar\omega$.

Acoustic deformation potential scattering rate is
\begin{equation}
 \Gamma^{ac} (E) = A_1 g(E)
\end{equation}

\begin{figure}[p]
\centering
\includegraphics[width=0.9\linewidth]{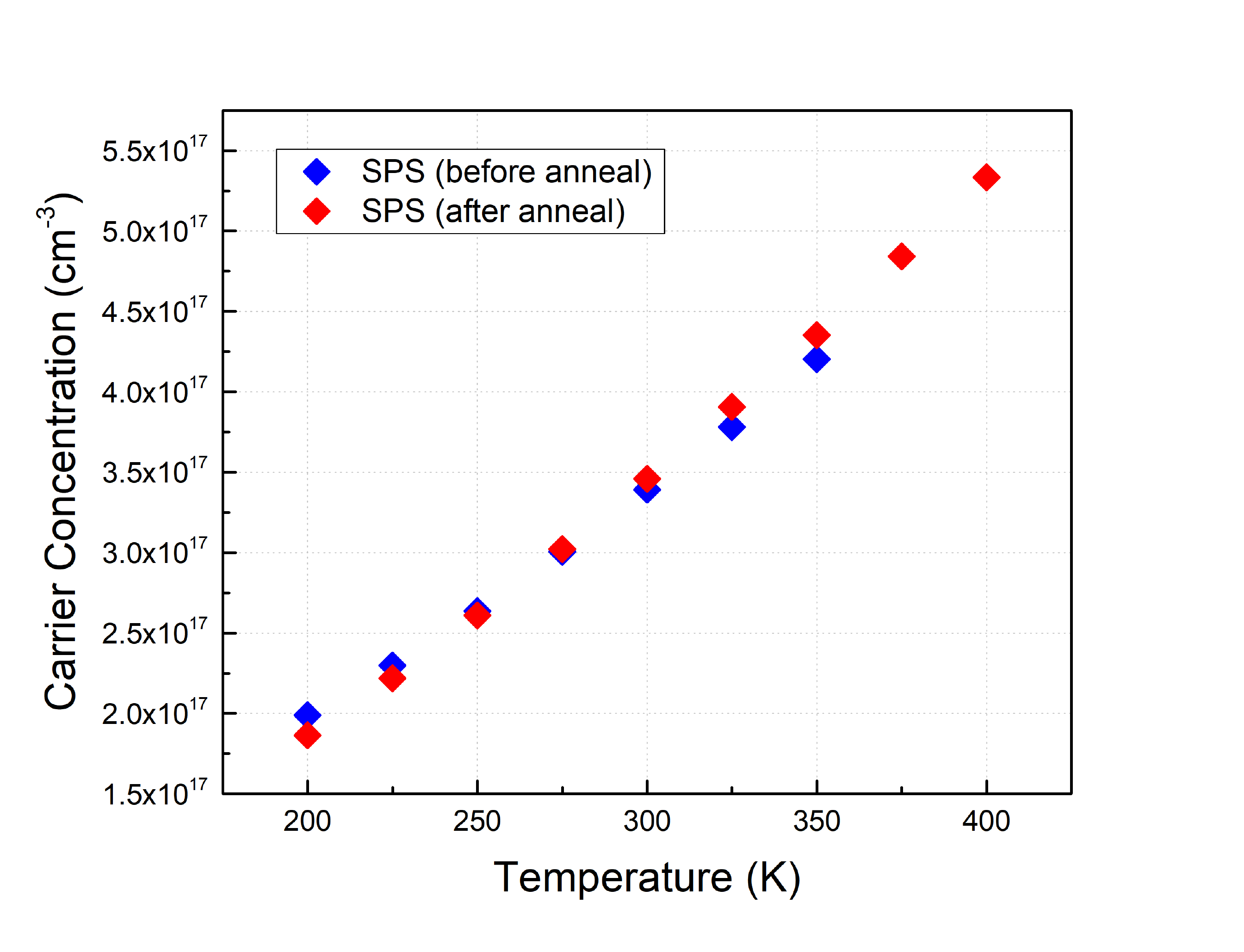}
\caption{The net carrier concentration obtained from the Hall coefficient measurements as a function of temperature for the samples before (blue diamonds) and after (red diamonds) 
annealing. The samples are found to be $n$-type.}
\label{fig:concentration}
\end{figure}

\begin{figure}[p]
\centering
\includegraphics[width=0.9\linewidth]{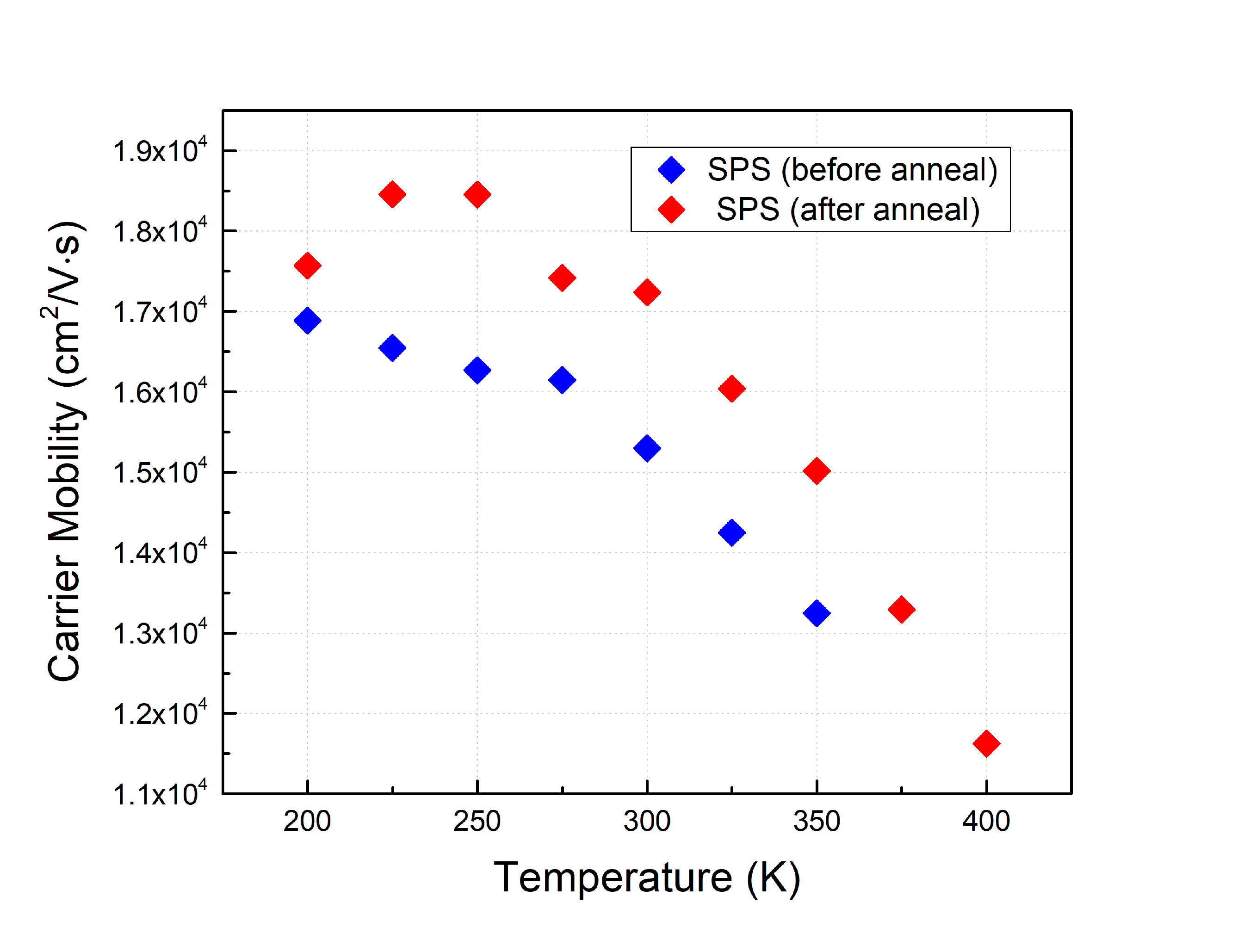}
\caption{The experimental carrier mobilities as a function of temperature for the samples before (blue diamonds) and after (red diamonds) annealing. The mobilities are impoved after annealing.
In both samples, the mobilities decrease with temperature.}
\label{fig:mobility}
\end{figure}

Ionized impurity scattering rate is
\begin{equation}
 \Gamma^{imp} (E) = A_2 n_{C} T E^{-3/2}
\end{equation}
where $n_C$ is the net carrier concentration obtained from the Hall coefficient measurements (see Fig.~\ref{fig:concentration})
\begin{equation}
n_C(T) = n_0 \exp(-T_d/T)
\end{equation}
where $T_d = 450.9$ K and $n_0 = 16.01\cdot10^{17}$ cm$^{-3}$. 

Polar optical scattering rate is 
$$ \Gamma^{pop} (E) = A_3\frac{\hbar\omega}{v_g} \left[ n_{BE}\sqrt{1+\frac{\hbar\omega}{E}} - n_{BE}\frac{\hbar\omega}{E}\sinh^{-1}\left(\sqrt{\frac{E}{\hbar\omega}}\right) + \right.$$
\begin{equation}
\left.+ (n_{BE}+1)\sqrt{1 - \frac{\hbar\omega}{E}} + (n_{BE}+1)\frac{\hbar\omega}{E}\sinh^{-1}\left(\sqrt{\frac{E}{\hbar\omega}-1}\right)\right]
\end{equation}
where $n_{BE}$ is the Bose-Einstein distribution function, $v_g$ is the group velocity. The first two terms represent the polar-optical absorption while the last two terms 
describe the emission. 

\begin{figure}[p]
\centering
\subfigure{\includegraphics[width=0.65\linewidth]{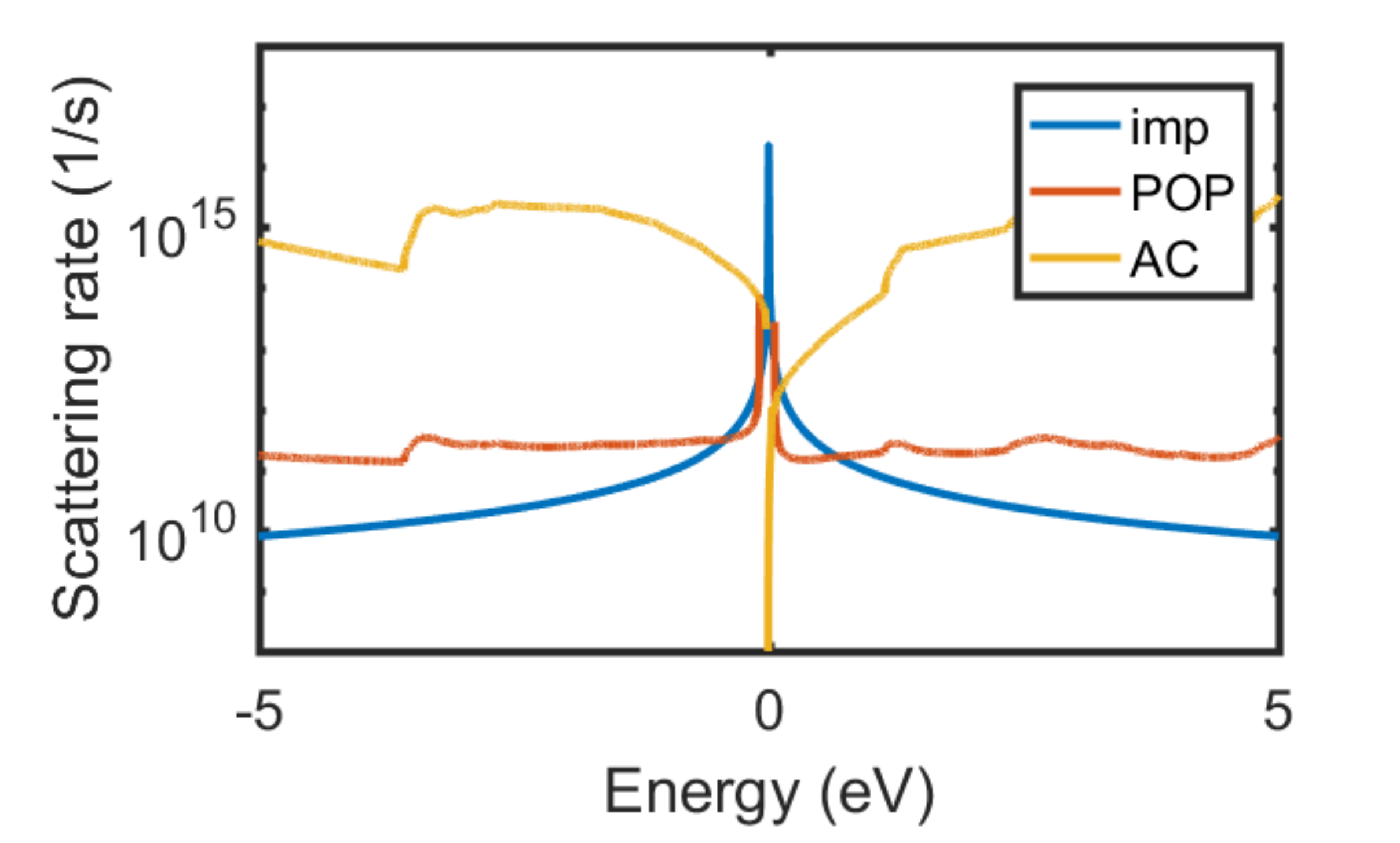}}
\hfill
\subfigure{\includegraphics[width=0.65\linewidth]{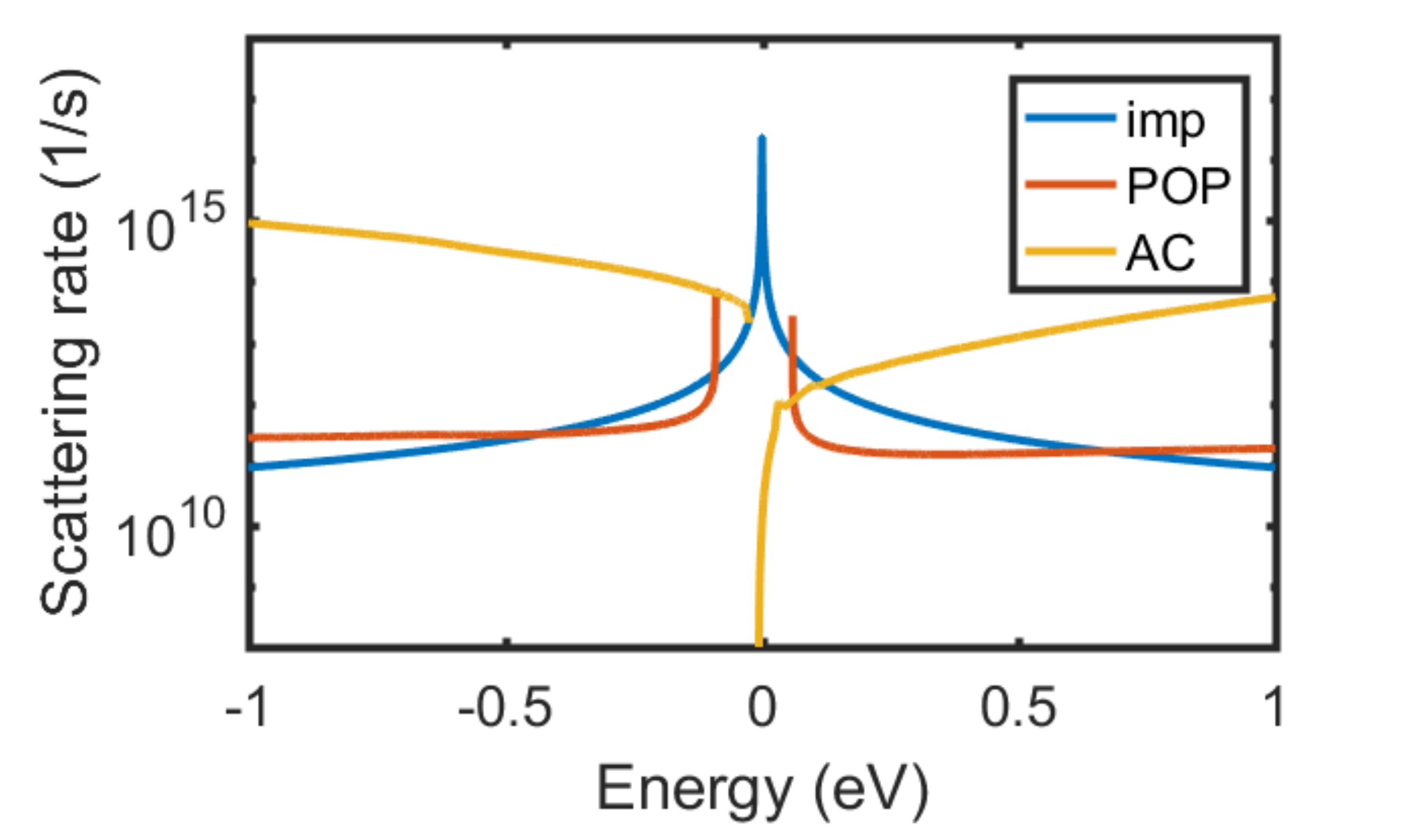}}
\caption{Acoustic deformation potential (yellow curves), polar optical (maroon curves) and charged impurity (blue curves) scattering rates obtained from the fitting of experimental electrical 
conductivities in the samples before (top panel) and after (bottom panel) annealing.}
\label{fig:reltimes}
\end{figure}

The energy dependent scattering rates obtained from the fitting to experimental electrical conductivity for the samples before and after annealing are shown in Fig.~\ref{fig:reltimes}. 

\subsection{Thermal conductivity measurements.}
\begin{figure}[p]
\centering
\includegraphics[width=1.00\linewidth]{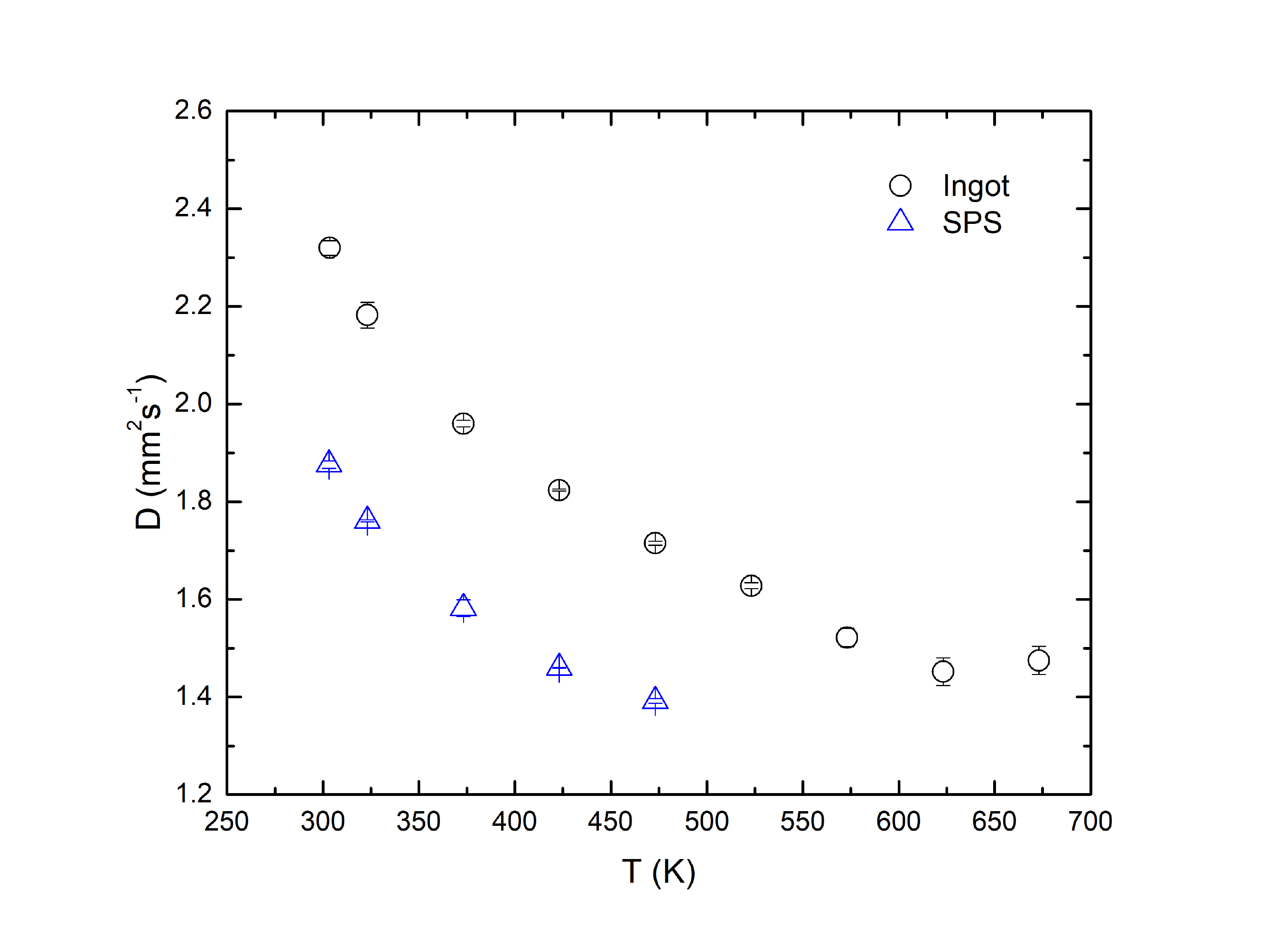}
\caption{The temperature-dependent thermal diffusivity of HgTe ingot and SPS samples. The thermal diffusivity decreases after the SPS process, and both of ingot and SPS samples’ 
thermal diffusivity reduce with increased temperature. }
\label{fig:diffusivity}
\end{figure}

To obtain the thermal conductivity $\kappa$, we use the following formula
\begin{equation}
 \kappa(T) = \rho c_p(T) D(T)
\end{equation}
where $\rho$ is the measured density of a sample, $D(T)$ is the measured thermal diffusivity and $c_p(T)$ is the theoretical specific heat capacity. 
The measured thermal diffusivity for the original ingot sample and the sample after the SPS is shown in Fig.~\ref{fig:diffusivity}. 
The ingot sample has an excess of Te atoms, and a lower density, $\rho=$7.82 $\pm$ 0.04 g/cm$^{3}$, comparing to $\rho =$ 7.98 $\pm$ 0.17 g/cm$^{3}$ after the SPS. 
The thermal diffusivity is higher for the ingot samples (black circles) than in the SPS samples (blue triangles), but does not change after the annealing of the SPS
sample. The theoretical heat capacity is 
\begin{equation}
c_p = c_v + V\frac{\alpha^2}{\beta_T}T
\label{eq:Cp}
\end{equation}
where $\alpha$ is the coefficient of thermal expansion, $\beta_T$ is the isothermal diffusivity, $c_V$ can be found from the Debye model
\begin{equation}
c_v = 9 N_A k_B \left(\frac{T}{T_D}\right)^3 \int_0^{x_D}dx \frac{x^4 e^x}{(e^x-1)^2} 
\end{equation}
where $T_D$ = 140 K is the Debye temperature. For the second term in Eq.~\ref{eq:Cp}, we use the experimental values from Ref.~\cite{Glazov:1996} and get the following expression 
for the specific heat
\begin{equation}
 c_{p}(T) = c_{V}(T) + 1.01\cdot10^{-2}T
\end{equation}
The obtained heat capacity $c_{p}(T)$ linearly changes from 0.158 JK$^{-1}$g$^{-1}$ at T = 250 K to 0.171 JK$^{-1}$g$^{-1}$ at T = 700 K
\subsection{Isotopic scattering for phonons.}
In this work, we perform \textit{ab initio} calculations solving the Boltzmann Transport Equation (BTE). The algorithm we use is described in details in Ref.~\cite{Fugallo:2013}. 
Apart from the intrinsic three-phonon scattering processes, we include the isotopic disorder scattering processes with rates given by 
\begin{equation}
 P^{iso}_{\mathbf{q}j} = \frac{\pi}{2N_{\mathbf{q}}} \omega_{\mathbf{q}j}\omega_{\mathbf{q'}j'}\delta(\hbar\omega_{\mathbf{q}j}-\hbar\omega_{\mathbf{q'}j'}) \left[n_{\mathbf{q}j}n_{\mathbf{q'}j'} + \frac{n_{\mathbf{q}j}+n_{\mathbf{q'}j'}}{2}\right]
 \sum_{s}g_{2}^{s} \left|\sum_{\alpha} z_{\mathbf{q}j}^{s\alpha}z_{\mathbf{q}'j'}^{s\alpha}\right|^2
\end{equation}
where $\mathbf{q}$ - phonon wave vector, $j$ - phonon branch index, $\omega_{\mathbf{q}j}$ - frequency of phonon $(\mathbf{q},j)$, $n_{\mathbf{q}j}$ - Bose-Einstein distribution function, $\alpha$ - Cartesian coordinate, $s$ - atom type,
$z_{\mathbf{q}j}^{s\alpha}$ - phonon eigenmode, $g_{2}^{s}$ - isotopic fluctuation parameter
\begin{equation}
 g_{2}^{s} = \frac{\sum_i c_i M_i^2 - \left(\sum_i c_i M_i\right)^2}{\left(\sum_i c_i M_i\right)^2}
\end{equation}
We use the natural isotopic composition of Hg and Te as summarized in Table~\ref{tab:iso}. The resulting isotopic fluctuation parameters are $g_{2}^{s} = 6.5\cdot10^{-5}$ for Hg and 
$g_{2}^{s} = 28.4\cdot10^{-5}$ for Te.

\begin{table}[h]
        \centering
\begin{tabular}{lcccr}
\hline
               $M_{Hg}$, amu & $\%$ & &$M_{Te}$, amu & $\%$ \\
\hline     
            195.966  &  0.15 & & 119.904 & 0.09 \\
            197.967  &  9.97 & & 121.903 & 2.55 \\
            198.968  &  16.87& & 122.904 & 0.89 \\
            199.968  &  23.10& & 123.903 & 4.74 \\
            200.970  &  13.18& & 124.904 & 7.07 \\
            201.971  &  29.86& & 125.903 & 18.84\\
            203.973  &  6.87 & & 127.904 & 31.74\\
                        &    & & 129.906 & 34.08\\
\hline
\end{tabular} 
\caption{\label{tab:iso} List of natural isotopes of Hg and Te.}
\end{table}

\subsection{Accumulated thermal conductivity.}

\begin{figure}[p]
\vspace{5cm}
\centering
\includegraphics[width=0.80\linewidth]{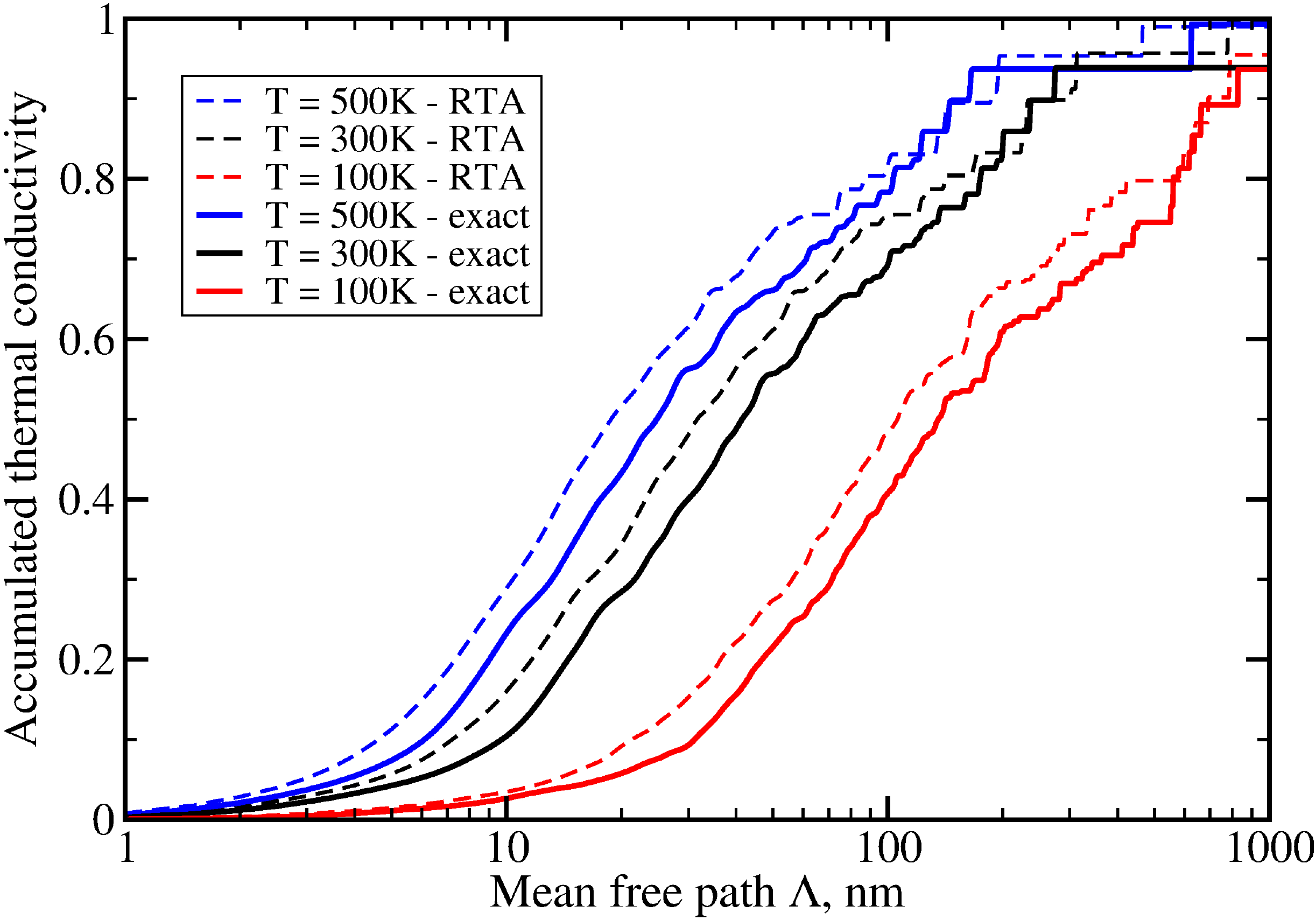}
\caption{Accumulated thermal conductivity calculated within the RTA (dashed lines) and from the exact solution of the BTE (solid lines).}
\label{fig:accumulated}
\end{figure}

The lattice thermal conductivity can be written as
\begin{equation}
\kappa_L = \frac{1}{k_B T^2 V_{cell} N_q} \sum_{\nu} n_{\nu} (1+n_{\nu}) \omega_{\nu}^2 c_{\nu} F_{\nu}
\end{equation}
where $V_{cell}$ is the unit cell volume, $\nu = \{\mathbf{q},j\}$, $c_{\nu}$ is the group velocity, $F_{\nu}$ is the linear deviation of the out-of-equilibrium phonon distribution 
$n^{out}_{\nu}$ from its equilibrium value $n_{\nu}$
\begin{equation}
n^{out}_{\nu} = n_{\nu} - \mathbf{F}_{\nu} \cdot \nabla T \frac{\partial n_{\nu}}{\partial T}
\end{equation}
It can be found from the solution of the Boltzmann Trasport Equation. In the relaxation time approximation (RTA) $F_{\nu}^{RTA} = \Lambda_{\nu}^{RTA} = \tau_{\nu} c_{\nu}$.
In the exact solution it plays a role of a vectorial mean free-path dispacement. To find a scalar mean-free path $\Lambda_{\nu}^{exact}$, one needs to project it onto velocity
direction
\begin{equation}
 \Lambda_{\nu}^{exact} = \frac{\mathbf{F}_{\nu} \cdot \mathbf{c_{\nu}}}{|\mathbf{c_{\nu}}|}
\end{equation}
The lattice thermal conductivity can be rewritten as a function of one single variable $\Lambda$ as
\begin{equation}
\kappa_L = \sum_{\nu} \kappa_L(\Lambda_{\nu}) = \int d\Lambda \kappa_L^{acc}(\Lambda)
\end{equation}
where the accumulated thermal conductivity is defined as 
\begin{equation}
\kappa_L^{acc}(\Lambda)= \sum_{\nu} \kappa_L(\Lambda) \delta(\Lambda-\Lambda_{\nu})
\end{equation}
In Fig.~\ref{fig:accumulated} we show the difference in the accumulated thermal conductivities in the two approaches discussed above. As one can see, the mean free path distribution 
in the exact approach is shifted toward the longer values.


\end{document}